\input amstex
\loadbold
\documentstyle{amsppt}
\pagewidth{32pc}
\pageheight{45pc}
\mag=1200
\baselineskip=15 pt
\NoBlackBoxes
\TagsOnRight

\def\gap{\vskip 0.1in\noindent}
\def\ref#1#2#3#4#5#6{#1, {\it #2,} #3 {\bf #4} (#5), #6.}

\def\aron {1} 
\def\avs {2} 
\def\brfs {3} 
\def\car {4} 
\def\cala {5} 
\def\cfs {6} 
\def\cyc {7} 
\def\dasi {8} 
\def\dasii {9} 
\def\deis {10} 
\def\del {11} 
\def\dss {12} 
\def\dono {13} 
\def\furs {14} 
\def\gp {15} 
\def\goso {16} 
\def\gor {17} 
\def\grfs {18} 
\def\ght {19} 
\def\hesj {20} 
\def\ish {21} 
\def\jito {22} 
\def\jlprl {23} 
\def\jli {24} 
\def\js {25} 
\def\kap {26} 
\def\katz {27} 
\def\kis {28} 
\def\kls {29} 
\def\kot {30} 
\def\kots {31} 
\def\kotfmv {32} 
\def\lastz {33} 
\def\lath {34} 
\def\lapaii {35} 
\def\lsii {36} 
\def\lesa {37} 
\def\pas {38} 
\def\pcmp {39} 
\def\pea {40} 
\def\rue {41} 
\def\sch {42} 
\def\ssg {43} 
\def\scmp {44} 
\def\svan {45} 
\def\ssp {46} 
\def\sst {47} 
\def\thou {48} 

\topmatter
\title Eigenfunctions, Transfer Matrices, and Absolutely 
Continuous Spectrum of One-dimensional Schr\"odinger 
Operators
\endtitle
\rightheadtext{Eigenfunctions, Transfer Matrices, and A.C.~Spectrum}
\author Yoram Last and Barry Simon$^{*}$
\endauthor
\leftheadtext{Y.~Last and B.~Simon}
\affil Division of Physics, Mathematics, and Astronomy \\ 
California Institute of Technology \\ Pasadena, CA 91125
\endaffil
\thanks$^*$ This material is based upon work supported by the 
National Science Foundation under Grant No.~DMS-9401491. 
The Government has certain rights in this material.
\endthanks
\endtopmatter

\document
\vskip 0.3in

\flushpar{\bf \S 1. Introduction}
\vskip 0.1in

In this paper, we will primarily discuss one-dimensional discrete 
Schr\"odinger operators
$$
(hu)(n)=u(n+1)+u(n-1)+V(n)u(n) \tag 1.1D
$$
on $\ell^2(\Bbb Z)$ (and the half-line problem, $h_+$, on $\ell^2 
(\{n\in\Bbb Z\mid n >0\}) \equiv \ell^2 ({\Bbb Z}^+)$) with $u(0)=0$ 
boundary conditions.  We will also discuss the continuum analog 
$$
(Hu)(x)=-u''(x)+V(x)u(x) \tag 1.1C
$$
on $L^2 (\Bbb R)$ (and its half-line problem, $H_+$, on $L^2 (0,\infty)$ 
with $u(0)=0$ boundary conditions).

We will focus on a new approach to the absolutely continuous spectrum 
$\sigma_{\text{\rom{ac}}}(h)$ and, more generally, $\Sigma_{\text{\rom{ac}}}
(h)$, the essential support of the a.c.~part of the spectral measures.

What is new in our approach is that it relies on estimates on the transfer 
matrix, that is, the $2\times 2$ matrix $T_E (n,m)$ which takes 
$\binom{u(m+1)}{u(m)}$ to $\binom{u(n+1)}{u(n)}$ for solutions $u$ of 
$hu=Eu$ (in the continuum case use $\binom{u'(x)}{u(x)}$ instead of 
$\binom{u(x+1)}{u(x)}$). We let $T_E (n)\equiv T_E (n,0)$. For example, 
we will prove the following: 

\proclaim{Theorem 1.1} Let $h_+$ be the operator {\rom{(1.1D)}} on $\ell^2 
(\{n\in\Bbb Z \mid n>0\})$ with $u(0)=0$ boundary conditions. Let 
$$
S=\biggl\{ E\biggm| \varliminf_{L\to\infty} \frac{1}{L}\, \sum^L_{n=1} 
\| T_E (n)\|^{2} < \infty\biggr\}.
$$
Then $S$ is an essential support of the a.c.~part of the spectral measure 
for $h_+$ \rom(i.e., $S=\Sigma_{\text{\rom{ac}}}(h)$\rom) and $S$ has 
zero measure with respect to the singular part of the spectral measure.
\endproclaim

The behavior of the transfer matrix is a reflection of the behavior 
of eigenfunctions since $T$ is built out of eigenfunctions. Indeed, if 
$u$ and $w$ are any two linearly independent solutions of $hu=Eu$ 
normalized at $0$, then $\frac{1}{L}\sum^L_{n=1} \|T(n)\|^2$ and 
$\frac{1}{L}\sum^{L+1}_{n=1} [|u(n)|^2 + |w(n)|^2]$ are comparable and 
so Theorem~1.1 relates the a.c.~spectrum to the behavior of eigenfunctions. 

That there is a connection between eigenfunctions and a.c.~spectrum is not 
new. Gilbert-Pearson [\gp] related a.c.~spectrum to subordinate solutions. 
Typical is the following (actually due to [\kap]; see also [\jlprl,\jli]): 
Call a solution $u$ of $hu=Eu$ subordinate if and only if for any linearly 
independent solution $w$,  
$$
\sum^L_{n=1} |u(n)|^2 \bigg/ \sum^L_{n=1} |w(n)|^2 \to 0 \tag 1.2
$$
as $L\to \infty$. Let
$$
S_0 =\{E\mid \text{there is no subordinate solution}\}.
$$
Then $S_0$ is an essential support of the a.c.~part of the spectral 
measure for $h_+$ and $S_0$ has zero measure with respect to the 
singular part of the spectral measure. 

The Gilbert-Pearson theory provides one-half of the proof of Theorem~1.1. 
Indeed, we will show that $S\subset S_0$. The other direction is intimately 
related to some new eigenfunction estimates which we discuss in Section~2. 
Its relation to the theory of Browder, Berezinski, Garding, Gel'fand, and 
Kac is discussed in the appendix. 

Related to Theorem~1.1 is the following, which also relies on the 
eigenfunction estimate of Section~2: 

\proclaim{Theorem 1.2} Let $h_+$ be as in Theorem~{\rom{1.1}}. Let $m_j, 
k_j$ be arbitrary sequences in $\{n\in\Bbb Z\mid n>0\}$ and let  
$$
S_1 =\biggl\{E\biggm| \varliminf_{j\to\infty} 
\|T_E (m_j, k_j)\| <\infty\biggr\}.
$$
Then $S_1$ supports the a.c.~part of the spectral measure for $h_+$ in 
that $\rho_{\text{\rom{ac}}}(\Bbb R\backslash S_1)=0$.
\endproclaim

These two theorems allow us to recover virtually all the major abstract 
results proven in the past fifteen years on the a.c.~spectrum for ergodic 
Schr\"odinger operators with the exception of Kotani's results [\kot,\kotfmv] 
on $\{E\mid\gamma (E)=0\}$. More significantly, they establish new results 
and settle an important open problem. Among the results recovered via a 
new proof are the Ishii-Pastur theorem [\ish,\pas], Kotani's  support 
theorem [\kots], and the results of Simon-Spencer [\ssp]. 

In a companion paper with A.~Kiselev [\kls], we will use Theorems~1.1, 1.2 
and Theorem~1.3 below to analyze, recover, and extend results on decaying 
random potentials [\scmp,\dss,\del], sparse potentials [\pcmp,\pea], and 
$n^{-\alpha} (1>\alpha >\frac{3}{4})$ potentials [\kis]. 

Theorem~1.1 and Fatou's lemma immediately imply that if $Q$ is any subset 
of $\Bbb R$ and  
$$
\sup_{n} \int_Q \|T_E (n)\|^2 \, dE <\infty, \tag 1.3
$$
then $Q$ lies in the essential support of $d\rho_{\text{\rom{ac}}}$ (for 
Fatou's lemma and (1.3) show for a.e.~$E\in Q$ we have that $\varliminf 
\frac{1}{L} \sum^L_{n=1} \|T_E (n)\|^2 <\infty$) but (1.3) does not seem 
to eliminate the possibility of singular spectrum on $Q$ (on the set of 
Lebesgue measure zero where Fatou does not apply). In this regard, the 
following result, which is an extension of ideas of Carmona [\car], is 
of interest:

\proclaim{Theorem 1.3} Suppose that 
$$
\varliminf_{n\to\infty} \int^b_a \|T_E (n)\|^p\,dE <\infty
$$
for some $p>2$. Then the spectrum is purely absolutely continuous on 
$(a,b)$.
\endproclaim

It is interesting to compare Theorems~1.2 and 1.3. A priori, one might 
think there could be potentials so there exist $n_1 <m_1 <n_2 <m_2 
<\cdots$ (with the $m_j -n_j$ and $n_{j+1} -m_j$ growing very rapidly) 
so that $T_E (n)$ is bounded at the $m_j$ and unbounded at the $n_j$. 
While this could happen at a single $E$, by these two theorems it cannot 
happen for all $E$ in $(a,b)$. 

To describe our most important new result, we define

\definition{Definition} Let $V,W$ be bounded functions on $\{n\in\Bbb Z
\mid n>0\}$. We say that $W$ is a right limit of $V$ if and only if there 
exist $n_j \to\infty$ so that $V(n + n_j) \to W(n)$ as $j\to\infty$ for 
each fixed $n>0$. 
\enddefinition

Then we will prove from Theorem~1.1 and the eigenfuction expansion results 
of Section~2 that

\proclaim{Theorem 1.4} If \, $W$ is a right limit of $V$ and $\tilde h_+, 
h_+$ are the half-line Schr\"odinger operators associated to $W,V$ 
respectively, then $\Sigma_{\text{\rom{ac}}} (h_+)\subset 
\Sigma_{\text{\rom{ac}}}(\tilde h_+)$. 
\endproclaim

\remark{Remark} This result is particularly interesting because it is 
easy to see that $\sigma_{\text{\rom{ess}}} (\tilde h_+) \subset
\sigma_{\text{\rom{ess}}} (h_+)$ with the inclusion in the opposite 
direction. 
\endremark

Our proof of Theorem~1.4 depends on the shift to transfer matrices 
rather than eigenfunctions.

This theorem will have an important corollary:

\proclaim{Theorem 1.5} Let $W$ be an almost periodic function on 
$\Bbb Z$ \rom(resp.~$\Bbb R$\rom). Let $h$ \rom(resp.~$H$\rom) be the 
full-line operator given by {\rom{(1.1)}}. For each $W_\omega$ in the 
hull of $W$\!, let $h_\omega$ \rom(resp.~$H_\omega$\rom) be the 
corresponding operator. Then the a.c.~spectrum, indeed the essential 
support of the a.c.~spectrum, of $h_\omega$ is independent of $\omega$.
\endproclaim

\remark{Remarks} 1. The result holds more generally than almost 
periodic potentials.  It suffices that the underlying process be 
minimally ergodic.

2. We will also recover the Deift-Simon [\deis] result that the 
multiplicity of the a.c.~spectrum is $2$.

3. Following Pastur [\pas] and others (see [\cala,\cyc]), it is known 
that the spectrum and its components are a.e.~constant on the hull. In 
1982, Avron-Simon [\avs] proved that the spectrum is everywhere constant 
rather than a.e.~constant in the almost periodic case. Theorem~1.5 has 
been believed for a long time, but this is its first proof. It is known 
(see Jitomirskaya-Simon [\js]) that the s.c.~ and p.p.~components need 
{\it{not}} be everywhere constant. 
\endremark

In this paper, we will also obtain rigorous spectral results on the 
operator $h_+$ where  $V(n)=\lambda\cos(n^\beta)$, and $1<\beta$ is 
not an integer.

Theorem~1.5 is reminiscent of the invariance of the a.c.~spectrum under 
rank one perturbations for all couplings. This is no coincidence. In our 
development of Theorems~1.1--1.2, what distinguishes a.c.~spectrum from 
non-a.c.~spectrum is its invariance under boundary conditions.  

While the main focus of this paper is on the a.c.~spectrum and 
transfer matrices, we will say something about point spectrum also. In 
this introduction, we will focus on the discrete case with $V$ bounded. 
In [\sst], using constancy of the Wronskian,  Simon-Stolz proved

\proclaim{Theorem 1.6 ([\sst])} If $\sum^\infty_{n=1} \|T_E (n)\|^{-2}=
\infty$, then $hu=Eu$ has no solution which is $L^2$ at infinity.
\endproclaim

As we will see in Section~8, it can happen that $\sum^\infty_{n=1} 
\|T_E (n)\|^{-2} <\infty$ without there being a solution $L^2$ at infinity; 
indeed, without there even being a bounded solution, but $\sum^\infty_{n=1} 
\|T_E (n)\|^{-2} <\infty$ has one important consequence.  Call a solution 
$u$ of $hu=Eu$ strongly subordinate if for any linearly independent solution 
$w$ we have that 
$$
[u(n)^2 +u(n+1)^2] \big/ [w(n)^2 + w(n+1)^2] \to 0
$$
as $n\to\infty$. It is easy to see that any strongly subordinate solution 
is subordinate. We will prove that

\proclaim{Theorem 1.7} If $V$ is bounded and $\sum^\infty_{n=1} 
\|T_E (n)\|^{-2} <\infty$, then there is a strongly subordinate solution of 
$hu=Eu$. This solution, $u_\infty$, obeys the estimate
$$
\|u_\infty (n)\|^2 \leq \| T_E(n)\|^{-2} + \frac{\pi^2}{4} \, 
\| T_E(n)\|^2 \biggl(\, \sum^\infty_{m=n} \frac{1}{\|T_E(m)\|^2}\biggr)^2.
$$
In particular, if
$$
\sum ^\infty_{n=1} \biggl\{\|T_E (n)\|^2 \biggl(\,\sum^\infty_{m=n} 
\|T_E (m)\|^{-2} \biggr)^2 \biggr\} <\infty,
$$
then $hu=Eu$ has an $L^2$ solution.
\endproclaim

Theorem~1.7 is essentially an abstraction of a well-known argument of 
Ruelle [\rue]. We will use it in [\kls,\lsii] to prove point spectrum in 
certain models, including new and simplified proofs of the results of 
Simon [\scmp] and some of the results of Gordon [\gor]. 

The plan of this paper is as follows. In Section~2 we develop 
eigenfunction estimates. Their relation to the BGK eigenfunction 
expansions is discussed in the appendix which includes higher-dimensional 
results. In Section~3 we use the results of Section~2 and the 
Gilbert-Pearson theory to prove Theorems~1.1 and 1.2 and we will use 
Carmona's formula to prove Theorem~1.3. In Section~4 we recover and 
extend the Simon-Spencer [\ssp] results. In Section~5 we prove Theorem~1.4 
and in Section~6 we prove Theorem~1.5 and some other consequences of 
Theorem~1.4, including the Kotani support theorem. In Section~7 we discuss 
$\lambda\cos (n^\beta)$.  In Section~8 we prove Theorems~1.7 and 1.8.  

We would like to thank Bert Hof and Svetlana Jitomirskaya for useful 
discussions.  B.S.~would like to thank M.~Ben-Artzi for the hospitality 
of the Hebrew University where some of this work was done.  

\vskip 0.3in
\flushpar{\bf \S 2. Eigenfunction Estimates}

\vskip 0.1in

We consider half-line problems in this section. In the discrete case for 
fixed $V(n)$ and $z\in\Bbb C$, define $u_D (n), u_N (n)$ to be the 
solution of $hu=zu$ ($h$ given by (1.1D)) with boundary conditions 
$$\align
u_D (0)=0 & \qquad u_D (1) = 1 \\
u_N (0) = 1 & \qquad u_N (1) = 0.
\endalign
$$
We will use $X$ to denote $D$ or $N$ in formulas where either is valid, 
and $Y$ for the opposite condition. 

In the continuum case, $u_D, u_N$ obey $Hu=zu$ ($H$ given by (1.1C)) with 
boundary conditions 
$$\align
u_D (0) = 0 & \qquad u'_D(0) = 1 \\
u_N (0)=1 & \qquad u'_N (0) = 0.
\endalign
$$
Of course, $u$ is $z$-dependent and we will sometimes use 
$u(\,\cdot \, ; z)$. It is standard that $u(n;z)$, $u(x;z)$, and $u'(x;z)$ 
are entire functions of $z$ for real $x,n$. 

The solutions $u$ are related to the transfer matrix by
$$
T_E (n) = \pmatrix u_N (n+1) & u_D (n+1) \\
u_N (n) & u_D (n) 
\endpmatrix \tag 2.1D
$$
in the discrete case and
$$
T_E (x) =\pmatrix u'_N (x) & u'_D (x) \\
u_N (x) & u_D (x)
\endpmatrix \tag 2.1C
$$
in the continuum case.

For $z\in\Bbb C_+ =\{z\mid\text{Im } z >0\}$, there is a unique solution 
$L^2$ at $+\infty$ (for arbitrary $V$ in the discrete case and for $V$ 
which is limit point at infinity in the continuum case). Both it and 
its derivative (in the continuum case) are everywhere non-vanishing. In 
the continuum, we denote the solution by $\varphi^D_+ (x;z)$ if normalized 
by $\varphi^D_+ (0;z)=1$ and $\varphi^N_+ (x;z)$ if normalized by 
$(\varphi^N_+)'(0;z)=-1$, and in the discrete case $\varphi^D_+ (0;z)=1$, 
$\varphi^N_+ (1;z)=-1$. This normalization is chosen so that the Wronskian 
of $\varphi^X_+$ and $u_X$ is $+1$. 

The $m$-functions are defined by
$$
\varphi^X_+ (\,\cdot\,; z)=\pm u_Y (\,\cdot\,; z) + m_X (z) u_X 
(\,\cdot\,;z) \tag 2.2
$$
where we take the plus sign in case $X=D$ and minus in case $X=N$. (Noting 
that the Wronskian of $\varphi^X_+$ and $\varphi^Y_+$ is zero, we see 
that $m_X (z)m_Y (z) = -1$.)

It is well known (see, e.g., [\cala,\katz]) that the $m$-functions are 
Herglotz (i.e., analytic with $\text{Im }m>0$ on $\Bbb C_+$) and that the 
measures 
$$
d\rho^X (E) =\lim_{\epsilon\downarrow 0}\,\frac{1}{\pi}\, \text{Im }m_X 
(E + i\epsilon)\, dE \tag 2.3
$$
are spectral measures for the operator $H_X$ ($h$ or $H$ with appropriate 
boundary conditions; i.e., in the continuum case $H_{D,N}$ are defined on 
$L^2 (0,\infty)$ with $u(0)$ or $u'(0)$ boundary conditions, and in the 
discrete case $H_D$ (resp.~$H_N$) is defined on $\ell^2 (\Bbb Z_+)$ 
(resp.~$\ell^2 (\{2,3,\dots,\})$) with $u(0)=0$ (resp. $u(1)=0$) boundary 
conditions). That is, $H_X$ is unitarily equivalent to multiplication by $E$ 
on $L^2 (\Bbb R, d\rho^X (E))$. Note that in (2.3) (and similarly (2.8) 
below), the limit is intended in the weak sense, that is, holds when 
smeared in $E$ with continuous functions of compact support.

In the discrete case and in the continuum case with $X=N$, we have
$$
\int \frac{d\rho^X (E)}{|E|+1} < \infty \tag 2.4a
$$
and
$$
m_X (z) = \int \frac{d\rho^X (E)}{E - z}. \tag 2.4b
$$
In the continuum case with $X=D$, we only have
$$
\int \frac{d\rho^D (E)}{E^2 + 1} <\infty \tag 2.5a
$$
and a Herglotz representation
$$
m_D (z)=a_0 + \int \biggl(\frac{1}{E - z} - \frac{E}{1+E^2}\biggr)\,
d\rho^D (E) \tag 2.5b
$$
for a suitable real constant $a_0$.

We are heading toward a proof of the following theorems:

\proclaim{Theorem 2.1D} In the discrete case, for any $V$ and $n$,
$$
\int |u_X (n;E)|^2 \,d\rho^X (E) = 1. \tag 2.6D
$$
\endproclaim

\proclaim{Theorem 2.1C} In the continuum case for any $V\geq 0$ and all $x$,
$$
\int \frac{|u_X (x;E)|^2}{E+1} \, d\rho^X (E) \leq \frac{1}{2} \,
(1\mp e^{-2|x|}) \tag 2.6C(a)
$$
where $\mp$ correspond to $X=D/N$. Moreover, for a universal constant $C$, 
we have that for all $x$ 
$$
\int \frac{\bigl[\int^{x+1}_{x-1} |u'_X (y;E)|^2 \,dy\bigr]}{(E+1)^2} \, 
d\rho^X  (E) \leq C . \tag 2.6C(b)
$$
\endproclaim

\remark{Remarks} 1. In (2.6C(b)), if $x<1$, interpret $x-1$ as $0$.

2. Obviously, $V\geq 0$ can be replaced by $V\geq c$ for any $c$ if 
$(E +1)^{-1}$ in (2.6C) is replaced by $(E -c+1)^{-1}$. The proof shows 
that as long as $-V \leq \alpha (-\Delta) + \beta$ for some $\alpha < 1$, 
estimates similar to (2.6C) hold (with $(E +1)^{-1}$ replaced by 
$(E + |\beta| + (1-\alpha)^{-1}$) and the $\frac{1}{2}$ (resp.~$1$) in 
the inequality replaced by $\frac{1}{2} (1-\alpha)^{-1}$ 
(resp.~$(1-\alpha)^{-1}$). Thus, the result allows any $V$ whose 
negative part is uniformly locally $L^1$. 
\endremark

\smallskip

As a preliminary we note that

\proclaim{Lemma 2.2} {\rom{(a)}} $\epsilon^2 |m(E + i\epsilon)|\to 0$ as 
$\epsilon \downarrow 0$ uniformly for $E$ in compact subsets of $\Bbb R$.

{\rom{(b)}} $\epsilon |\text{\rom{Re }} m(E +i\epsilon)|\to 0$ as $\epsilon 
\downarrow 0$ and is uniformly bounded for $E$ in compacts.
\endproclaim

\demo{Proof} (a) is a direct consequence of (2.4/2.5). (b) follows from 
those formulas and the dominated convergence theorem. \qed
\enddemo

The resolvent, $(H_X -z)^{-1}$, of the operator $H_X$ has a continuous 
integral kernel (in the continuum case). In general, this kernel 
$G_X (x,y;z)$ has the form 
$$
G_X (x,y;z)=u_X (x_< ;z) \varphi^X_+ (x_>; z) \tag 2.7
$$
where $x_< =\min (x,y)$, $x_> =\max (x,y)$. This formula is easy to 
verify and shows that $G$ is continuous. 

\proclaim{Theorem 2.3}
$$
\lim\limits_{\epsilon\downarrow 0}\,\frac{1}{\pi}\, \text{\rom{Im}}\, G_X 
(x,x; E+i\epsilon)\, dE = |u_X (x,E)|^2 \, d\rho^X (E). \tag 2.8
$$
\endproclaim

\demo{Proof}  By (2.7) and (2.2),
$$
G_X (x,x; E + i\epsilon)=\pm u_Y (x, E +i\epsilon) u_X (x; E +i\epsilon) 
+m_X (E +i\epsilon) u_X (x; E +i\epsilon)^2.
$$ 
Since $u_{X,Y}$ are entire and real for $z$ real, we have that 
$\lim _{\epsilon\downarrow 0} \,\text{Im}\, u_Y (x , E +i\epsilon) u_X 
(x, E +i\epsilon) =0$. Similarly, $u_X (x; E + i  \epsilon)^2 = u_X (x; E)^2 
+ i\epsilon a(x; E) + O(\epsilon^2)$ where $u^2_X$ and $a(x)$ are real. Thus, 
$$
\text{Im }[m_X (E +i\epsilon) u_X (x; E +i\epsilon)^2] =  \boxed{1} + 
\boxed{2} + \boxed{3}
$$
with
$$
\boxed{1} = u_X (x;E)^2 \text{ Im }m_X (E +i\epsilon)\to 
\pi |u_X (x,E)|^2 \,d\rho^X (E)
$$
by (2.3) and
$$
\boxed{2}=\epsilon a(x; E)\text{ Re }m(E +i\epsilon)\to 0
$$
by Lemma~2.2(b) and
$$
\boxed{3} =\text{Im}[O(\epsilon^2)m(E +i\epsilon)]\to 0
$$
by Lemma~2.2(a). Thus, (2.8) is proven. \qed
\enddemo

\remark{Remarks} 1. (2.8) is essentially a version of the spectral theorem. 
We will discuss this further in the appendix.

2.  The same method shows more generally that 
$$
\lim\limits_{\epsilon\downarrow 0} \frac{1}{\pi}\, G_X (x,y; E +i\epsilon) =
u_X (x,E) u_X (y,E)\, d\rho^X (E). \tag 2.8$^\prime$
$$

3.  (2.8/2.8$^\prime$) are not new; they are implicit, for example, in 
Section~II.3 of Levitan-Sargsjan [\lesa].
\endremark

\demo{Proof of Theorem {\rom{2.1D}}} (2.8) says that $|u|^2 \,d\rho$ is 
the spectral measure for $H_X$ with vector $\delta_n$. Thus, $\int 
|u(n;E)|^2\,d\rho^X (E)= (\delta_n, \delta_n) =1$. \qed 
\enddemo

\demo{Proof of Theorem {\rom{2.1C}}} $G_D (x,x;z)$ is analytic in 
$\Bbb C\backslash [0,\infty)$ and goes to zero as $|z| \to \infty$. 
It follows that
$$
G_X (x,x; -1)=\int \frac{|u(x,E)|^2\, d\rho^X (E)}{E +1}\,.
$$
But since $V\geq 0$, $(H_X +1)^{-1}\leq (H^{(0)}_X +1)^{-1}$ where 
$H^{(0)}_X$ is the operator when $V=0$. Thus, 
$$
G_X (x,x;-1) \leq G^{(0)}_X (x,x;-1)=\frac{1}{2}\, (1 \pm e^{-2|x|})
$$
by the method of images formulas for $G^{(0)}$. This proves (2.6C(a)).

To prove (2.6C(b)) where $x\geq 2$, pick $g$ a $C^\infty$ function with 
$0\leq g\leq 1$, $g$ supported on $[-2,2]$, and $g\equiv 1$ on $[-1, 1]$. 
Let $f(y)=g(y-x)$. Then 
$$\align
\int^{x+1}_{x-1} (u')^2\, dy & \leq \int f^2 (u')^2\, dy \\
& = -\int f^2 u''u \, dy -\frac{1}{2}\int (u^2)' (f^2)' \, dy \\
& = \int f(E -V) u^2\, dy + \frac{1}{2} \int (f^2)'' u^2\, dy \\
& \leq C(1+|E|) \int^{x+2}_{x-2} u^2\, dy.
\endalign
$$
Thus,  (2.6(b)) for $x\geq 2$ follows from (2.6(a)).

A similar calculation works for $x=1$. Explicitly, pick $f$ which is 
supported on $[0,3)$ and $f\equiv 1$ on $[0,2]$. Because $u(0)u'(0)=0$, 
the above calculations still show that 
$$\align
\int^2_0 |u(y)'|^2\, dy & \leq \int f(E-V) u^2\, dy + \frac{1}{2} \int 
(f^2)'' u^2 \, dy \\
&  \leq C(1+|E|) \int^3_0 u^2 \, dy.
\endalign
$$
(2.6(b)) for $x=1$ and $x\geq 2$ imply the result for all $x$. \qed
\enddemo

\remark{Remark} If $V$ is uniformly locally $L^2$, one can show that 
(2.6C(b)) holds without the need for integrating over $y$.
\endremark

\vskip 0.3in
\flushpar{\bf \S 3. Criteria for A.C.~Spectrum}
\vskip 0.1in

Our main goal in this section is to prove Theorems~1.1 and 1.2 as well 
as a continuum analog of Theorem~1.2.  We begin with an estimate based 
on the Gilbert-Pearson theory and then apply the bounds of Section~2. We 
will then provide a new proof of the Pastur-Ishii theorem. Finally, we 
present a condition for purely a.c.~spectrum. 

Fix $V$ and $E$. For each $\theta\in [0,\pi)$, let $\Phi_\theta$ be the 
vector formed from the solution with $(\sin\theta, \cos\theta)$ boundary 
conditions at $0$, that is,
$$
\Phi_\theta (\,\cdot\,)=T_E (\,\cdot\,) \binom{\sin\theta}{\cos\theta} 
\tag 3.1a
$$
and let $\Psi_\theta$ be $\Phi_{\pi/2 +\theta}$, that is,
$$
\Psi_\theta (\,\cdot\,)=T_E (\,\cdot\,) \binom{\cos\theta}{-\sin\theta}. 
\tag 3.1b
$$
Define $u_\theta, v_\theta$ by $\Phi_\theta (n) = 
\binom{u_\theta (n+1)}{u_\theta(n)}$, $\Psi_\theta(n) = 
\binom{v_\theta (n+1)}{v_\theta (n)}$.

The Wronskian of  $u$ and $v$ is constant, that is, $\langle \Phi, J\Psi
\rangle =1$ with $J=\left(\smallmatrix 0&1 \\ -1&0\endsmallmatrix\right)$. 
It follows by the Cauchy-Schwarz inequality that
$$
\| \Phi_\theta (n)\| \, \|\Psi_\theta (n)\| \geq 1. \tag 3.2
$$
Clearly, $\| \Psi_\theta (\,\cdot\,)\| \leq \|T_E (\,\cdot\,)\|$ by (3.1b). 
Let us use the symbol $\frac{1}{L} \int^L_0 \cdot \, dx$ for the integral 
in the continuum case and for the sum $\frac{1}{L} \sum^L_{n=1}\cdot\,$ 
in the discrete case. Then 
$$
\frac{1}{L} \int^L_0 \|\Psi_\theta (x)\|^2\, dx \leq \frac{1}{L} \int^L_0 
\|T_E (x) \|^2 \, dx. \tag 3.3
$$
By (3.2),
$$\align
1 &\leq \biggl( \frac{1}{L}\int^L_0 \|\Phi_\theta (x)\| \, 
\|\Psi_\theta (x)\| \, dx\biggr)^2 \\
& \leq\biggl( \frac{1}{L} \int^L_0 \|\Phi_\theta (x)\|^2 \biggr) 
\biggl( \frac{1}{L} \int^L_0 \|\Psi_\theta (x)\|^2 \, dx\biggr). \tag 3.4
\endalign
$$
(3.3) and (3.4) immediately imply

\proclaim{Lemma 3.1}
$$
\frac{\int^L_0 \| \Psi_\theta (x)\|^2 \, dx}
{\int^L_0 \| \Phi_\theta (x)\|^2 \, dx} \leq
\biggl( \frac{1}{L} \int^L_0 \| T_E (x)\|^2 \, dx \biggr)^2 
\tag 3.5
$$
\endproclaim

Recall the definitions of Gilbert-Pearson. A solution $u_\theta$ is 
called subordinate if and only if
$$
\lim\limits_{x\to\infty} \, 
\frac{\int^L_0 | u_\theta (x)|^2\, dx}{\int^L_0 | v_\theta (x) |^2 \, dx} 
= 0. \tag 3.6 
$$

To use (3.6), we must deal with the fact that $\Phi, \Psi$ are not quite 
the same as $u,v$.  In the discrete case, we have that 
$$
\sum^{L+1}_{n=1} | v_\theta (n)|^2 \leq \sum^L_{n=1} 
\| \Psi_\theta (n)\|^2
$$
while
$$\align
\sum^{L+1}_{n=1} |u_\theta (n) |^2 &\geq \frac12 \sum^L_{n=1} 
(|u_\theta (n)|^2 + |u_\theta (n+1)|^2) \\
&\geq \frac12 \sum^L_{n=1} \| \Phi_\theta (n)\|^2
\endalign
$$
so returning to $\int^L_0 \,\cdot\, dx$ notation for the sum
$$
\frac{\int^{L+1}_0 |v_\theta (x)|^2 \, dx}
{\int^{L+1}_0 |u_\theta (x)|^2 \, dx} \leq 
\frac{2 \int^L_0 \| \Psi_\theta (x)\|^2\, dx}
{\int^L_0 \| \Phi_\theta (x)\|^2 \, dx} 
$$
so
$$
\frac{\int^{L+1}_0 |v_\theta (x)|^2\, dx}
{\int^{L+1}_0 |u_\theta (x)|^2 \, dx} \leq 2 
\biggl( \frac{1}{L} \int^L_0 \|T_E (x) \|^2\, dx  \biggr)^2. 
\tag 3.7D
$$

In the continuum case, one can mimic the proof of Theorem~2.1C to 
see that if $V \geq 0$, then 
$$
\int^L_0 (u_\theta (x)^2 + u'_\theta (x)^2)\, dx \leq C (1+|E|) \int^{L+1}_0 
u_\theta (x)^2\, dx.
$$
Thus, 
$$\align
\frac{\int^{L+1}_0 v^2_\theta (x)\, dx}{\int^{L+1}_0 u^2_\theta (x)\, dx} & 
\leq C(1+|E|) \, \frac{\int^{L+1}_0 \| \Psi_\theta (x)\|^2 \, dx}
{\int^L_0 \| \Phi_\theta (x)\|^2 \, dx} \\
&\leq C (1+|E|) \biggl( \frac{1}{L} \int^L_0 \| \Psi_\theta (x) \|^2 \, 
dx \biggr) \biggl( \frac{1}{L} \int^{L+1}_0 \| \Psi_\theta (x)\|^2 \, 
dx \biggr) \\
&\leq \biggl( \frac{(L+1)}{L}\biggr)^2 C(1+|E|) \biggl( \frac{1}{L+1} 
\int^{L+1}_0 \|T_E (x) \|^2\, dx \biggr)^2. \tag 3.7C
\endalign
$$
(3.6) and (3.7) imply that

\proclaim{Theorem 3.2} If $H$ has a subordinate solution at energy $E$, 
then
$$
\lim\limits_{L\to \infty} \, \frac{1}{L} \int ^L_0 \| T_E (x)\|^2\, dx 
= \infty. \tag 3.8
$$
\endproclaim

Let $Q = \{E\mid H \text{ has a subordinate solution at energy } E\}$ and 
let $S_0 = \Bbb R \backslash Q$. Recall $S$, the set of Theorem~1.1, is 
given by 
$$
S= \biggl\{E \biggm| \varliminf\limits_{L\to\infty} \frac{1}{L} \int ^L_0 
\| T_E (x)\|^2 \, dx <\infty \biggr\}.
$$
Theorem~3.2 says that $Q\subset \Bbb R \backslash S$ so $S\subset S_0$.  
Gilbert-Pearson have shown that $S_0$ is the essential support of the 
a.c.~part, $\mu_{\text{\rom{ac}}}$, of the spectral measure of $H_+$. 
Thus, $S\subset S_0$ implies that if $A\subset S$ and $|A| > 0$, then 
$\mu_{\text{\rom{ac}}}(A) >0$. Theorem~1.1  thus follows from 

\proclaim{Proposition 3.3} For a.e.~$E$ w.r.t.~ $\mu_{\text{\rom{ac}}}$, 
we  have that $E\in S$.
\endproclaim

\demo{Proof} In terms of the measures $d\rho^X$ of Section~2, let $d\mu(E) 
= \min (d\rho^D, d\rho^N)$ in the discrete case and $d\mu =(1+E^2)^{-1} 
\min (d\rho^D, d\rho^N)$ in the continuum case, where the min is defined 
viz.
$$
\min(\mu_1, \mu_2)(S) = \inf \Sb A,B \\ S\subset A \cup B \endSb 
\, \{\mu_1 (A) + \mu_2 (B) \}.
$$
Since the singular parts of $d\rho^D$ and $d\rho^N$ are disjoint and the 
a.c.~parts are mutually equivalent (see, e.g., [\svan]),  $d\mu$ is 
equivalent to the a.c.~part of the spectral measure for $H_+$. By (2.1) and 
(2.6), we have that for each $n$, 
$$
\int d\mu (E) \| T_E (n)\|^2 \leq 4 \tag 3.9D
$$
in the discrete case and for each $x_0 \geq 1$,
$$
\int d\mu (E) \int ^{x_0 +1}_{x_0 -1} \| T_E (x)\|^2 \leq C \tag 3.9C
$$
in the continuum case. Here $C$ is a universal constant. It follows that
$$
\int d\mu (E) G_L (E) \leq C,  \tag 3.10
$$
where
$$
G_L (E) =\frac{1}{L} \sum^L_{n=1} \|T_E (n)\|^2
$$
in the discrete case and
$$
G_L (E) =\frac{1}{Q(L)}\, \int ^L_0 \| T_E (x)\|^2\, dx,
$$
where $Q(L)$ is the smallest even integer less than $L$ (so $L/Q(L) 
\to 1$ as $L\to\infty$). 

By (3.10) and Fatou's lemma, $\int d\mu (E) \varliminf G_L (E) <\infty$, 
so $\varliminf G_L (E) <\infty$ a.e. w.r.t.~$d\mu$, that is, $E\in S$ for 
a.e.~$E$ w.r.t.~to $d\mu$. \qed
\enddemo

\remark{Remark}  An immediate consequence of Theorem~1.1 is that if 
$V_\omega$ is an ergodic family of potentials and the Lyapunov exponent 
$\gamma (E)>0$ on a Borel set $T\subset\Bbb R$, then for a.e.~$\omega$, 
$\Sigma_{\text{\rom{ac}}} (H_\omega) \cap T =\emptyset$. For by Fubini's 
theorem for a.e.~$\omega$, for a.e.~$E\in T$, we have $\lim \frac{1}{n} 
\ln \|T(n)\| >0$ so that a fortiori, $\lim \frac{1}{L}\sum^L_{n=1} 
\|T(n)\|^2 = \infty$ and thus, for a.e.~$\omega$, $S\cap T$ has zero 
Lebesgue measure. This result is the celebrated Ishii-Pastur theorem 
[\ish,\pas,\cala,\cyc]. Note that our proof is more direct than the one 
that goes through the construction of exponentially decaying eigenfunctions. 
\endremark

To prove Theorem~1.2 (and also Theorem~1.4), we need to extend (3.9) from 
$T(n)$ to $T(n,m)$. As in that equation, $d\mu$ is the $\min$ of $d\rho^N$ 
and $d\rho^D$ which is an a.c.~measure for $h_+$:

\proclaim{Theorem 3.4D}  For any $n,m$, $\int \|T_E (n,m)\| d\mu (E) 
\leq 4$.
\endproclaim

\demo{Proof}  We have that $\| T(n,m)\|\leq \|T(n,0)\| \, \|T(0,m)\| = 
\| T(n,0)\| \, \|T(m,0)\|$, so by the Schwarz inequality, 
$$
\int \| T_E (n,m)\| \, d\mu \leq \biggl( \int \| T_E (n,0)\|^2 \, 
d\mu\biggr)^{1/2} \biggl(\int \| T_E (m,0)\|^2 \, d\mu\biggr)^{1/2} 
\leq 4
$$
by (3.9). \qed
\enddemo

An immediate consequence of this theorem and Fatou's lemma is

\proclaim{Theorem 3.5D ($\equiv$ Theorem 1.2)} Let $m_j, k_j$ be 
arbitrary sequences in $\{n\in\Bbb Z\mid n >0\}$. Then for a.e.~$E$ in 
the a.c.~part of the spectral measure for $h_+$, we have that 
$\varliminf _{j\to\infty} \|T_E (m_j, k_j)\|  < \infty$. 
\endproclaim

The continuum versions of these results are straightforward analogs 
following the above proof using (3.9C). Here $d\mu (E)=(1+E^2)^{-1} 
\min (d\rho^D (E), d\rho^N (E))$.

\proclaim{Theorem 3.4C} For each $x_0, y_0$ and a universal constant $C$,
$$
\int d\mu(E) \biggl[ \, \int ^{x_0 +1}_{x_0 -1} dx 
\int ^{y_0 +1}_{y_0 -1} dy \, \| T_E (x,y)\| \biggr] < C.
$$
\endproclaim

\proclaim{Theorem 3.5C}  Let $x_j, y_j$ be arbitrary sequences in 
$\{x\in\Bbb R \mid x > 0\}$. Then for a.e.~$E$ in the a.c.~part of the 
spectral measure for $H_+$, we have that 
$$
\varliminf\limits_{j\to\infty} \int ^{x_j +1}_{x_j -1} dx 
\int^{y_j +1}_{y_j -1} dy \, \| T_E (x,y)\| < \infty.
$$
\endproclaim

We will need the following variant of these ideas in Section~5:

\proclaim{Theorem 3.6D} In the discrete case,
$$
\int d\mu(E) \biggl( \frac{1}{L} \sum^{n+L}_{m=n+1} \|T(m,n)\|^2 
\biggr)^{1/2} \leq 4. \tag 3.11
$$
\endproclaim

\demo{Proof} Since $\|T(m,n)\| \leq \|T(m)\| \, \|T(n)\|$, we have that 
$(\frac{1}{L} \sum^{n+L}_{m=n+1} \|T(m,n)\|^2) \leq \mathbreak \|T(n)\| 
(\frac{1}{L} \sum^{n+L}_{m=n+1} \|T(m)\|^2 )^{1/2}$, so (3.11) follows 
from (3.9D) and the Schwarz inequality. \qed 
\enddemo

In the same way, we get

\proclaim{Theorem 3.6C} In the continuum case for a universal constant $C$,
$$
\int d\mu (E) \biggl( \, \int ^{x_0 +1}_{x_0 -1} dx \, \frac{1}{Q(L)} 
\int ^{x+L}_{x} \| T_E (x,y)\|^2 \, dy \biggr)^{1/2} \leq C.
$$
\endproclaim

\vskip 0.1in

\centerline{* \qquad * \qquad *}

\vskip 0.1in

\proclaim{Theorem 3.7 ($=$ Theorem 1.3)} Suppose that for some $x_n \to 
\infty$,
$$
\lim_{n\to\infty} \int ^b_a \|T_E (x_n)\|^p\, dE <\infty
$$
for some $p>2$. Then for any boundary condition at zero, the spectral 
measure is purely absolutely continuous on $(a,b)$. More generally, if $W$ 
is an arbitrary function on $(-\infty, \infty)$ so that 
\roster
\item"\rom{(i)}" $W=V$ on $(0,\infty)$
\item"\rom{(ii)}" $W$ is limit point at both $-\infty$ and $\infty$.
\endroster
Then $H= -\frac{d^2}{dx^2}+W$ has purely a.c.~spectrum on $(a,b)$.
\endproclaim

\demo{Proof}  Fix a boundary condition $\theta$ at zero and let $u_\theta 
=(\cos (\theta), \sin(\theta))$. For any $x$, let 
$$
d\mu^\theta_x (E) =\pi^{-1} dE \big/ \|T_E (x) u_\theta \|^2. 
\tag 3.12
$$
Then Carmona [\car] proves that as $x\to\infty$,
$$
d\mu^\theta_x \to d\mu^\theta, \tag 3.13
$$
the spectral measure for boundary condition $\theta$ (the convergence in 
(3.13) is in the vague sense, i.e., it holds after smearing with continuous 
functions in $E$).  Since $T$ is unimodular, $\|T^{-1}\| = \|T\|$ so 
$\| Tu_\theta \| \geq \| T\|^{-1} \|u_\theta \|$ and thus, $d\mu^\theta_x 
(E) = F^\theta_x (E)\, dE$ with 
$$
|F_x (E)| \leq \|T_E (x)\|^2. \tag 3.14
$$
For the whole-line problem, Carmona proves a result similar to (3.12/3.13), 
but in (3.12) $\|T_E (x) u_\theta \|$ is replaced by $\| T_E u_{\theta (E)}\|$  
with $\theta (E)$ dependent on $E$ (and $x$) but (3.14) still holds. The result 
now follows from the next lemma. \qed 
\enddemo

\proclaim{Lemma 3.8} Let $f_n (\lambda)$ be a sequence of functions on $(a,b)
\subset \Bbb R$ so that for some $q>1$,
$$
\int f_n (\lambda)^q \, d\lambda \leq C
$$
uniformly in $n$. Suppose that $f_n (\lambda) \, d\lambda $ converge to 
a measure $d\mu (\lambda)$ weakly. Then $d\mu$ is purely absolutely continuous.  
\endproclaim

\demo{Proof} The ball of radius $C$ in $L^q$ is compact in the weak-* topology, 
so there exists a subsequence $f_{n(i)}$ and $f_\infty \in L^p$ so that 
$\int f_{n(i)}(\lambda) g(\lambda) \, d\lambda \to \int f_\infty (\lambda) 
g(\lambda) \, d\mu(\lambda)$ for all $g\in L^{q'}$ with $q'$ dual to $p$. 
Thus, $d\mu =f_\infty \, d\lambda $ is absolutely continuous. \qed 
\enddemo

\vskip 0.1in
\centerline{* \qquad  * \qquad *}
\vskip 0.1in

We end this section with two remarks that shed some light on the earlier 
theorems in this section. The first concerns an explicit relationship between 
the $m$-function and the basic average $\frac{1}{L} \int^L_0 \|T_E (x)\|^2 \, 
dx$ which is connected with Lemma~3.1:

\proclaim{Proposition 3.9} We have for any $\theta$ that
$$
\text{\rom{Im }} m_\theta \biggl( E+i \,\frac{1}{L}\biggr) \leq 
\left(5 + \sqrt{24}\right) \biggl[ \frac{1}{L} \sum^{L+1}_{n=0} 
\|T_E (n)\|^2 \biggr]
$$
where $\|T_E (0)\|$ is short for $1$.
\endproclaim

\demo{Proof} Let $u_1$ be the solution with $\theta$ boundary conditions 
normalized at $n=1$ and $u_2$ the solution with complementary $(\frac{\pi}
{2} -\theta)$ boundary conditions. Then Jitomirskaya-Last [\jlprl,\jli] 
prove that if $\|f\|_L = (\sum^L_{n=1} f(n)|^2)^{1/2}$ and $\epsilon (L)$ 
is defined by 
$$
\|u_1\|_L \, \|u_2\|_L =(2\epsilon)^{-1}, \tag 3.15
$$
then
$$
|m(E+i\epsilon)| \leq \left(5+\sqrt{24}\, \right) \, 
\frac{\|u_2\|_L}{\|u_1\|_L}\, .
$$

If $L$ is odd, let $f=(u_1 (1), u_1 (2), \dots, u_1 (L-1))$ and let 
$g=(u_2 (2), -u_2 (1), u_2 (4), -u_2 (3), \mathbreak \dots, -u_2 (L-1))$. 
Then constancy of the Wronskian implies that $\langle f,g\rangle =\frac{L}{2}$, 
so by the Schwarz inequality,
$$
\frac{L-1}{2} \leq \|u_1\|_L \, \|u_2 \|_L = \frac{1}{2\epsilon (L)}\, . 
\tag 3.16
$$
For $L$ even, the inequality holds with $\frac{L}{2}$, so a fortiori, 
(3.16) holds. Thus, $\frac{1}{L}\geq \epsilon(L+1)$. Since 
$\text{Im }m (E+i\epsilon)/ \epsilon$ is monotone increasing as 
$\epsilon$ decreases, 
$$\align
\frac{\text{Im }m(E+iL^{-1})}{L^{-1}} &\leq 
\frac{\text{Im } m(E+i\epsilon(L+1))}{\epsilon(L+1)} \\
&\leq \frac{|m (E+i\epsilon)|}{\epsilon} \leq 
\frac{5+\sqrt{24}}{\epsilon} \, \frac{\|u_2\|_{L+1}}{\|u_1\|_{L+1}}\, .
\endalign
$$
By (3.15), $(\epsilon \|u_1\|_{L+1})^{-1} \leq 2\|u_2\|_{L+1}$, so
$$
\frac{\text{Im }m(E+iL^{-1})}{L^{-1}} \leq 2 \left( 5+\sqrt{24} \right) 
\|u_2\|^2_{L+1}\,.
$$

Now $|u_2 (n)|^2 + |u_2 (n+1)|^2 \leq \|T(n)\|^2$, so
$$
\sum^{L+1}_{n=0} \|T(n)\|^2 \geq |u(0)|^2 + 
|u(L+2)|^2 + 2 \|u_2\|^2_{L+1},
$$
proving that $2\|u_2\|^2_{L+1} \leq \sum^{L+1}_{n=0} \|T(n)\|^2$ and 
the claimed inequality. \qed
\enddemo

The second result concerns the fact that $\varliminf \frac{1}{L} 
\sum^L_{n=1} \|T(n)\|^2 <\infty$ says nothing about upper bounds. We claim 
that this sum cannot grow too fast, at least for a.e.~$E$ w.r.t.~$d
\mu_{\text{\rom{ac}}}$. 

\proclaim{Theorem 3.10} Fix $\delta > 0$. For a.e.~$E$ 
w.r.t.~$d\mu_{\text{\rom{ac}}}$, we have that for any $L\geq 2$,
$$
\biggl( \frac{1}{L} \sum^L_{n=1} \|T_E (n)\|^2 \biggr) \leq C_E 
(\log L)^{1+\delta}.
$$
\endproclaim

\remark{Remarks} 1. $(\log L)^{1+\delta}$ can be replaced by any increasing 
function $f(n)$ with $\sum f(2^n)^{-1} <\infty$, for example, $(\log L)(\log 
(\log L))^{1+\delta}$.

2. If we replace $\| T_E (n)\|$ by $\| u(n; E)\|$, this result holds for $d\mu 
(E)$ rather than just for $d\mu_{\text{\rom{ac}}}(E)$.
\endremark

\demo{Proof} Let $g_k (E) =2^{-k} \sum^{2^k}_{n=1} \|T_E (n)\|^2$. Then by 
(3.9D), $\int g_k (E)\, d\mu_{\text{\rom{ac}}} (E) \leq 4$ so \linebreak 
$\sum^\infty_{k=1} k^{-1-\delta} g_k (E) \in L^1 (d\mu_{\text{\rom{ac}}})$, 
which, in particular, implies that
$$
g_k (E) \leq C_E k^{1+\delta} \tag 3.17
$$
for a.e.~$E$ w.r.t.~$d\mu_{\text{\rom{ac}}}$.

Let $2^{k-1} \leq L\leq 2^k$. Then
$$
L^{-1} \sum^L_{n=1} \|T_E (n)\|^2 \leq 2^{-k-1} \sum^{2^k}_{n=1} 
\|T_E (n) \|^2 \leq 2 g_k (E),
$$
so (3.17) completes the proof. \qed
\enddemo

\vskip 0.3in
\flushpar {\bf {\S 4. Barriers and A.C.~Spectrum}}
\vskip 0.1in

Theorem~1.2, which we proved in Section~3, is ideal for showing that 
barriers can prevent a.c.~spectrum, an idea originally developed by 
Simon-Spencer [\ssp]. In this section, we will explain how to recover 
their results using Theorem~1.2. Our techniques here allow one to go 
further since they can handle the case where $V$ goes to zero. We will 
illustrate this at the end of this section. A more thorough analysis of 
this case will be made in a forthcoming paper [\lsii]. As the simplest 
example of the strategy, we recover

\proclaim{Theorem 4.1 ([\ssp])} Let $h_+$ be a Jacobi matrix on $\ell^2 
({\Bbb Z}^+)$. Suppose $\varlimsup |V(n)| = \infty$. Then $h_+$ has no 
a.c.~spectrum. 
\endproclaim

\demo{Proof} Pick $n_j$ so $|V(n_j)|\to\infty$. Then
$$
T_E (n_j, n_j -1) = \pmatrix E - V(n_j) & -1 \\ 1 & 0 \endpmatrix,
$$
so for all $E$, $\|T_E (n_j, n_j -1)\|\to\infty$ as $j\to \infty$. By 
Theorem~1.2, the a.c.~spectrum must be empty. \qed
\enddemo

To recover some of the other results of [\ssp], we need bounds that show 
if $E$ is in the middle of a gap of size $2\delta$, then the transfer 
matrix over a length $L$ has an a priori bound that grows as $L\to\infty$ 
in a way independent of the potential. We could obtain this using 
Combes-Thomas estimates with explicit constants (as in [\ssp]) or using 
the periodic potential methods of [\lath,\lapaii], but we will instead 
use the idea of approximate eigenfunctions. Our simple estimates can be 
viewed as a quantitative version of an idea of Sch'nol [\sch]. Basically, 
we will see that any solution is found to grow exponentially at a 
pre-assigned rate in some direction.

\proclaim{Theorem 4.2} Suppose $h$ is a one-dimensional operator of the form 
{\rom{(1.1D)}} on a subset $D$ of $\Bbb Z$ with $D_n \equiv \{-n, - n+1, 
\dots, n-1, n\}\subset D$. Suppose that there is an operator $B$ on $\ell^2 
(D')$ for some $D'\subset \Bbb Z$ with $D_n \subset D'$ so that 
\roster
\item"\rom{(i)}" $\text{\rom{spec}}(B) \cap (E-\delta, E+\delta) = \emptyset$.
\item"\rom{(ii)}" $Bu = hu$ if $u$ vanishes outside $D_n$.
\endroster
Then
\roster
\item"\rom{(a)}" Any solution of $hu=Eu$ obeys
$$
|u(\ell)|^2 + |u(-\ell)^2| \geq \delta^2 (1+\delta^2)^{\ell-1} |u(0)|^2 \quad 
\text{for }\ell=1,2, \dots, n+1 \tag 4.1
$$
and
$$
|u(\ell)|^2 + |u(-\ell)|^2 \geq \delta^2 (1+\delta^2)^{\ell-2} 
(|u(0)|^2 + |u(1)|^2 + |u(-1)|^2) \quad \text{for }\ell = 
2,3,\dots, n+1. \tag 4.2
$$
\item"\rom{(b)}" For any vector $\varphi\in \Bbb R^2$,
$$
\| T(\ell,0) \varphi\|^2 + \| T(-\ell, 0)\varphi\|^2 \geq \delta^2 
(1+\delta^2)^{\ell-1} \| \varphi\|^2 \quad \text{for } \ell=
1,2,\dots, n. \tag 4.3 
$$
\item"\rom{(c)}" We have that
$$
\| T(-n, n)\| \geq \tfrac12 \delta^2 (1+\delta^2)^{n-1}. \tag 4.4
$$
\endroster
\endproclaim

\remark{Remarks} 1. One remarkable aspect of these estimates is that 
they (and their multidimensional case continuum analog) are independent 
of $V$\!.

2. The point, of course, is that since $\delta >0$, $(1+\delta^2)^\ell$ 
grows to infinity as $\ell\to\infty$. We show it is exponentially fast, 
but that is not needed.

3. While the estimates are elegant and explicit, it is likely the exponent 
is not optimal. For $\delta$ small, $(1+\delta^2)^n \simeq \exp(n \log 
(1+\delta^2))\sim \exp(n\delta^2)$. One would expect that $\| T(n, -n)\| 
\sim \exp(2\delta n)$ for $\delta$ small (and fixed) and $n$ large. 
\endremark

\demo{Proof} Let $\chi_j$ be the characteristic function of 
$\{-j, \dots, j\}$. Then
$$
((h-E)(\chi_j u)) (\ell) = -\delta_{j,\ell+1} u(\ell+1) - 
\delta_{j, -\ell-1} u(-\ell-1)
$$
so if we define
$$
a_j \equiv |u(j)|^2 + |u(-j)|^2 \quad \text{for } j=1,2,\dots
$$
and
$$
a_0 \equiv |u(0)|^2
$$
we have that
$$
\| (H-E) (\chi_j u) \|^2 = a_{j+1}. \tag 4.5
$$
Clearly,
$$
\| \chi_j u\|^2 = \sum^j_{k=0} a_k. \tag 4.6
$$
But by hypothesis (i), (ii), if $j=0,1,\dots, n$,
$$
\| (H-E)\chi_j u\|^2 = \| (B-E)\chi_j u \|^2 \geq 
\delta^2 \| \chi_j u \|^2. \tag 4.7
$$
(4.5), (4.6), and (4.7) imply that
$$
\delta^2 \biggl( \, \sum^j_{k=0} a_k \biggr) \leq a_{j+1} \tag 4.8
$$
for $j=0,1,2,\dots, n$.

It follows inductively that for $\ell=1,2,\dots$
$$
a_\ell \geq \delta^2 (1+\delta^2)^{\ell-1} a_0 \tag 4.9
$$
for (4.9) holds for $\ell=1$ (by 4.8), and if (4.9) holds for $a_1, \dots, 
a_j$, then by (4.8), 
$$
a_{j+1} \geq \delta^2 \biggl( 1+\sum^j_{k=1} \delta^2 (1+\delta^2)^{k-1} 
\biggr) a_0 = \delta^2 (1+\delta^2)^j a_0.
$$
(4.9)  is precisely (4.1).

A virtually identical inductive argument proves (4.2).  (4.3) follows from 
(4.1) and its translate:
$$
|u(\ell + 1)|^2 + |u(-\ell+1)|^2 \geq \delta^2 (1+\delta^2)^{\ell-1} 
|u(1)|^2; \qquad \ell=1,\dots, n.
$$

To prove (4.4), let $\alpha = \frac12 \delta^2 (1+\delta^2)^{n-1}$ so that 
(4.3) becomes
$$
\| T(n, 0) \varphi\|^2 + \|T(-n,0)\varphi\|^2 \geq 2\alpha 
\| \varphi\|^2. \tag 4.10
$$
If $\alpha \leq 1$, (4.4) is trivial so suppose that $\alpha >1$. Picking 
any unit vector $\varphi$, we conclude that
$$
\| T(n,0)\|^2 \geq \alpha \qquad \text{or} \qquad \| T(-n,0)\|^2 \geq \alpha.
$$

Suppose the former. Since $T(n,0)$ is unimodular, we can find a unit 
vector $\varphi_0$ so that $\| T(n,0) \varphi_0\| = \|T(n,0)\|^{-1}$. Thus,
$$
\| T(n,0) \varphi_0 \|^2 \leq \frac{1}{\alpha} \, \| \varphi_0 \|^2 \leq 
\alpha \|\varphi_0 \|^2
$$
because we are supposing that $\alpha >1$. Thus, by (4.10),
$$
\| T(-n, 0) \varphi_0 \|^2 \geq \alpha \| \varphi_0 \|^2 \geq \alpha^2 
\| T(n,0)\varphi_0 \|^2.
$$
It follows that
$$
\| T(n, -n)\|^2 = \| T(-n, n)\|^2 \geq \alpha^2
$$
which is (4.4). \qed
\enddemo

Once we have Theorem~4.2, we immediately conclude by Theorem~1.2 that

\proclaim{Theorem 4.3} Suppose that $h$ has the form {\rom{(1.1D)}} on 
${\Bbb Z}^+$ so that there exist $x_n \geq n$ and $W_n$ on $\tilde D_n 
\supset \{ x_n - n, \dots, x_n + n\}$ so that
\roster
\item"\rom{(i)}" $(\alpha,\beta) \cap \text{\rom{spec}} (-\frac{d^2}
{dx^2}+ W_n ) = \emptyset$ for some boundary conditions on $\tilde D_n$. 
\item"\rom{(ii)}" $W_n (j) = V(j)$ for $j\in \{x_n -n, \dots, x_n +n\}$. 
\endroster
Then $(\alpha, \beta)$ is disjoint from the a.c.~spectrum of $h$.
\endproclaim

\demo{Proof} Fix $E\in (\alpha, \beta)$. Then by Theorem~4.2,
$$
\lim_{n\to\infty} \| T_E (x_n -n, x_n + n)\| = \infty.
$$
It follows by Theorem~1.2 that $(\alpha, \beta)$ is disjoint from the 
a.c.~spectrum. \qed
\enddemo

With this result, one can recover the theorems in [\ssp] that depend on 
gaps in the spectrum.

Before leaving the subject of Theorem~4.2, we note that (4.1) has a 
continuum, higher-dimensional analog.

\proclaim{Theorem 4.4} For any $K>0$ and dimension $\nu$, there exists a 
universal constant $C_\nu (K)$ depending only on $\nu$ and $K$ so that 
if $V$ is in the local Kato class and there exists an operator $B$ on $L^2 
(\Bbb R^\nu)$ so that 
\roster
\item"\rom{(i)}" $Bu = (-\Delta + V)u$, all $u\in C^\infty_0 (D_n)$ where 
$D_n = \{ x\mid |x| \leq n+1 \}$ 
\item"\rom{(ii)}" $\sigma (B) \cap (E-\delta, E+\delta)=\emptyset$
\item"\rom{(iii)}" $\| V\chi_{\{x\mid \, |x| \leq n+1\}} \| \leq K$ where 
the norm is the $K_\nu$ Kato class norm {\rom{[\cyc,\ssg]}} 
\item"\rom{(iv)}" $|E| \leq K$ 
\endroster
then any $L^2_{\text{\rom{loc}}}$ distributional solution of $(-\Delta + V)u 
= Eu$ in $D_n$ obeys
$$
\int_{j\leq |x|\leq j+1} |u(x)|^2\, dx \geq C_\nu (K)\delta^2 (1+C_\nu (K) 
\delta^2)^{j-2} \int_{|x| \leq 1} |u(x)|^2 \, dx \tag 4.11
$$
for $j=1,2,\dots, n$.
\endproclaim

\demo{Proof} Let $\chi_j$ be the characteristic function of $\{ x\mid |x| 
\leq j\}$. It is fairly easy to see one can construct a sequence, $f_j$, of 
$C^\infty$ functions on $\Bbb R^\nu$ so that 
$$\align
f_j \chi_j &= \chi_j \\
f_j \chi_{j+1} &= f_j
\endalign
$$
and
$$
\sup_j \| D^\alpha f_j \| \equiv d_\alpha < \infty \tag 4.12
$$
for each multi-index $\alpha$.

We claim that with $H=-\Delta + V$\!,
$$
\| (H-E) f_j u \|^2 \leq C_\nu (K)^{-1} \| (\chi_{j+1} - \chi_j) u\|^2. 
\tag 4.13
$$
Accepting this for a moment, we will prove (4.8). We have for $j\leq n-1$,
$$
\| (H-E) f_j u \|^2 = \| (B-E) f_j u\|^2 \geq \delta^2 \| f_j u\|^2 
\geq \delta^2 \| \chi_j u \|^2.
$$
Thus with $a_j = \| (\chi_{j+1} - \chi_j)u \|^2$, we see that
$$
C_\nu \delta^2 \biggl(\, \sum^j_1 a_\ell \biggr) \leq a_{j+1}
$$
so that as in the proof of Theorem~4.2,
$$
a_j \geq C_\nu \delta^2 (1+C_\nu \delta^2)^{j-2} a_1
$$
which is (4.11).

To prove (4.13), notice that
$$
(H-E) f_j u = (-\Delta f_j) u + 2(\nabla f_j) \, \cdot \, \nabla u
$$
so
$$
\| (H-E) f_j u \|^2 \leq 2 \| (-\Delta f_j) u\|^2 + 8 \| (\nabla f_j) 
\, \cdot \, \nabla u \|^2. \tag 4.14
$$
By Theorem~C.2.2 of [\ssg], we can bound $\| \nabla f \, \cdot\, 
\nabla u\|^2$ by a constant $C_1$ (depending on $K$) times 
$\| (\chi_{j-1} - \chi_j) u\|^2$ so by (4.14), we have the 
estimate (4.13). \qed 
\enddemo

\proclaim{Theorem 4.5} Fix $\alpha < \frac12$ and let 
$\{a_n\}^\infty_{n=1}$ be identically independently distributed random 
variables with distribution $\frac12 \chi_{[-1,1]} (x)\, dx$. Then there 
exists $N_1 < N_2 <\cdots $ so that for any $m_1, \dots, m_n, \dots \geq 0$ 
and a.e.~$\{a_n\}$ the potential on $\Bbb Z^+$:
$$
V(n) = \cases 0 & n\leq m_1 \\ 
(n-m_1)^{-\alpha} a_n & m_1 < n \leq m_1 + N_1 \\
0 & m_1 + N_1 < n \leq m_1 + N_1 + m_2 \\
\vdots & {} \\
(n-m_1 - N_1 - \cdots - m_j)^{-\alpha} a_n & m_1 + \cdots + N_{j-1} + 
m_j < n \leq m_1 +\cdots + N_j \\
0 & m_1 + \cdots + N_j < n \leq m_1 + \cdots + N_j + m_j 
\endcases
$$
has no a.c.~spectrum.
\endproclaim

\remark{Remark} The choice can be made so that by Theorem~1.6, there is no 
point spectrum, that is, so the spectrum is purely singular continuous.
\endremark

\demo{Proof} Let $\tilde T_E (0,n)$ be the transfer matrix for the 
power-decaying potential $n^{-\alpha} a_n$. By [\scmp], for a.e.~$\{a_n\}$ 
and a.e.~$E\in [-2,2]$, 
$$
\lim\limits_{n\to\infty} \|\tilde T_E (0,n)\| = \infty. \tag 4.15
$$
Let $A^{(1)}_\ell = \{ E\mid \inf_{n\geq \ell} \| \tilde T_E (0,n)\| 
\geq 1 \}$. By (4.15) $|[-2,2]\backslash A_\ell | \downarrow 0$ as $\ell 
\to \infty$ so we can pick $N_1$ so that $[-2,2] \backslash|A^{(1)}_{N_1}| 
\leq 2^{-1}$. Now inductively pick $N_j$ given $N_1, \dots, N_{j-1}$ so if
$$
A^{(j)}_\ell = \biggl\{ E \biggm| \inf\limits_{n\geq\ell} \| \tilde T_E 
(N_1 + \cdots + N_{j-1}, N_1 + \cdots + N_{j-1} + n)\| \geq n\biggr\}
$$
then $[-2,2]\backslash |A^{(j)}_{N_j}| \leq 2^{-j}$.

For this choice of $N_j$'s, the theorem holds since for a.e.~$E$, $E\in 
A^{(j)}_{N_j}$ for all large $j$ and thus for such $E$, $\| T_E (m_1 + 
\cdots + N_{j-1}+m_j, m_1 + \cdots + m_j + N_j) \| \geq j$. Theorem~3.5 
implies $\sigma_{\text{\rom{ac}}} = \emptyset$. \qed
\enddemo

[\kls] will have a much more effective analysis of this type of example.

\vskip 0.3in
\flushpar {\bf {\S 5. Semicontinuity of the A.C.~Spectrum}}
\vskip 0.1in

In this section, we will prove Theorem~1.4. Consider first the discrete 
case. Pick $n_j$ so $V(n+ n_j) \to W(n)$ as $j\to\infty$ for each $n$. 
Let $T_V$ (resp.~$T_W$) denote the transfer matrix for the Jacobi matrix 
with $V$ (resp.~$W$) along the diagonal. By Theorem~3.6D, 
$$
\int d\mu_V (E) \biggl( \frac{1}{L} \sum^{n_j + L}_{m=n_j + 1} 
\| T_V (m, n_j)\|^2 \biggr)^{1/2} \leq 4 \tag 5.1
$$
where $d\mu_V (E)$ is a measure equivalent to the a.c.~part of the 
spectral measure for $V$\!. Since $V(n+n_j)\to W(n)$ as $j\to\infty$, 
we have that
$$
T_V (n_j + m, n_j) \to T_W (m,0)
$$
so (5.1) implies that
$$
\int d\mu_V (E) \biggl( \frac{1}{L} \sum^L_{m=1} \| T_W (m,0)\|^2 
\biggr)^{1/2} \leq 4. \tag 5.2
$$

It follows by Fatou's lemma that for a.e.~$E$ with respect to 
$d\mu_V (E)$, we have 
$$
\varliminf \, \frac{1}{L} \sum^L_{m=1} \|T_W (m,0)\|^2 < \infty.
$$
Such $E$ are thus a.e.~in $\Sigma_{\text{\rom{ac}}} (h_0 +W)$, that is, 
$\Sigma_{\text{\rom{ac}}} (h_0 + V) \subset \Sigma_{\text{\rom{ac}}} 
(h_0 + W)$ as claimed. \qed

\smallskip
The proof in the continuum case is similar, except that we use 
Theorem~3.6C in place of Theorem~3.6D.

We note that the notion of right/left limits, which enters in Theorem~1.4, 
is in the spirit of the notion of limit class introduced by Davies-Simon 
[\dasii].

\vskip 0.3in
\flushpar {\bf {\S 6. Consequences of Semicontinuity of the A.C.~Spectrum}}
\vskip 0.1in

Let $(\Omega, T, \mu)$ be a metric ergodic process, that is, $T$ is a 
continuous invertible bijection from $\Omega \to \Omega$ with $\Omega$ a 
compact metric space (recall that any separable compact space is metrizable) 
and $\mu$ a probability measure with support $\mu = \Omega$. 

\definition{Definition} A point $\omega_0 \in \Omega$ is called right 
prototypical if and only if $\{T^n \omega_0 \mid n \geq 0\}$ is dense in 
$\Omega$, and left prototypical if and only if $\{T^n \omega_0 \mid 
n\leq 0\}$ is dense in $\Omega$. If $\omega_0$ is both left and right 
prototypical, we say it is prototypical.
\enddefinition

The ergodic theorem implies that a.e.~$\omega_0 \in \Omega$ is 
prototypical. Fix a continuous function $f \: \Omega \to \Bbb R$ and 
let $h_\omega$ on $\ell^2 (\Bbb Z)$ be defined by $(h_\omega u)(n) = 
u(n+1) + u(n-1) + f(T^n \omega)u(n)$. 

\proclaim{Theorem 6.1} The essential support of the a.c.~spectrum of 
$h_\omega$ is the same for all prototypical points and is of 
multiplicity $2$. Moreover, for any prototypical $\omega_0$ and any 
$\omega\in \Omega$, we have $\Sigma_{\text{\rom{ac}}} (h_{\omega_0}) 
\subset \Sigma_{\text{\rom{ac}}} (h_\omega)$.
\endproclaim

\demo{Proof} Let $h^\pm_\omega$ be the operators on $\ell^2 (\pm n 
\geq 1)$ with $u(0) =0$ boundary conditions. By general principles (see, 
e.g., Davies-Simon [\dasi]), the restriction of $h_\omega$ to its 
a.c.~subspace is unitarily equivalent to the restriction of  $h^+_\omega 
\oplus h^-_\omega$ to its a.c.~subspace. Thus, the theorem follows from
\roster
\item"\rom{(i)}" if $\omega_0$ is prototypical and is $\omega$ arbitrary, 
then $\Sigma_{\text{\rom{ac}}} (h^\pm_{\omega_0}) \subset 
\Sigma_{\text{\rom{ac}}} (h^\pm_\omega )$  
\item"\rom{(ii)}" for prototypical $\omega_0$, $\Sigma_{\text{\rom{ac}}} 
(h^+_{\omega_0}) = \Sigma_{\text{\rom{ac}}} (h^-_{\omega_0})$
\endroster
for (i) implies equality if both $\omega_0$ and $\omega$ are prototypical 
and (ii) implies multiplicity 2.

To prove (i), pick $n_j \to \infty$ so $T^{+n_j} \omega_0 \to \omega$. 
Then, since $V_{T^{+m}\omega_0} (n) = f (T^{n+m} \omega_0) = V_{\omega_0} 
(n+m)$, we have that $V_{\omega_0} ( \, \cdot\, + n_j) \to V_\omega 
(\, \cdot \,)$ so (i) follows from Theorem~1.4.

To prove (ii), let $\omega_0$ be prototypical and $\omega\neq \omega_0$, 
also prototypical. Pick $n_j \to \infty$ so $T^{+n_j} \omega_0 \to \omega$. 
Fix $L$ and use the fact that $\| T(n,m)\| = \| T(m,n)\|$ (since $T$ is 
unimodular) to note that 
$$\align
\frac{1}{L} \sum^L_{m=1} \| T_\omega (-1, -m) \|^2 &= \frac{1}{L} 
\sum^L_{m=1} \| T_\omega (-m, -1) \|^2 \\
&= \lim \frac{1}{L} \sum^L_{m=1} \| T_{\omega_0} (n_j - m, n_j -1)\|^2 
\endalign
$$
so by Theorem~3.6D,
$$
\int d\mu_{\omega_0} (E) \biggl(\frac{1}{L} \sum^L_{m=1} 
\| T_\omega (-1, -m)\|^2 \biggr) < 4
$$
where $d\mu_{\omega_0} (E)$ is an a.c.~measure for $h^+_{\omega_0}$. Thus, 
as in the last section, $\Sigma_{\text{\rom{ac}}} (h^-_\omega) \supset 
\Sigma_{\text{\rom{ac}}} (h^+_{\omega_0})$. By symmetry, (ii) holds. \qed 
\enddemo

\remark{Remark} That the typical a.c.~spectrum is of multiplicity 2 is a 
result of Deift-Simon [\deis] proven using Kotani theory. Our proof is 
different.

If $V$ is almost periodic, then every $\omega \in \Omega$ is prototypical. 
Thus, Theorem~6.1 implies Theorem~1.5. More generally, if $(T, \Omega, \mu)$ 
is minimal (or if it is strictly ergodic which implies minimal), then every 
$\omega \in \Omega$ is prototypical, and we see that the a.c.~spectrum is 
constant (rather than just a.e.~constant) on $\Omega$. 
\endremark

\example{Example} Consider the sequence $V _1, V_2, V_3, \dots $ given by
$$
0,1,0,0,0,1,1,0,1,1,0,0,0,0,0,1,0,1,0, \dots
$$
defined as follows. For two finite sequences of $0$'s and $1$'s of length 
$n$, say $w_1, \dots, w_n$, and $\tilde w_1, \dots, \tilde w_n$, say 
$w < \tilde w$, if and only if 
$$
w_1 = \tilde w_1, \dots, w_j = \tilde w_j, \qquad w_{j+1} < \tilde w_{j+1}.
$$
With this order, the sequences of length $n$ are well-ordered, for example,
$$
(0)<(1), \quad (0,0) < (0,1) < (1,0) < (1,1), \quad 
(0,0,0) < (0,0,1) < (0,1,0) < \cdots.
$$
$V$ is obtained by placing the two sequences of length 1 in order, then 
the four sequences of length 2, etc. Clearly,  $V$ is prototypical for a 
Bernoulli model. By Furstenberg's theorem, that model has no a.c.~spectrum, 
so $V$ is an explicit sequence for which we know that 
$\sigma_{\text{\rom{ac}}} (h_0 + V) = \emptyset$. 
\endexample

Another consequence of Theorem~6.1 is a new proof of the Kotani support 
theorem:

\proclaim{Theorem 6.2} Let $\Omega$ be the compact metric space of 
sequences $V_n$ with $|V_n|\leq a$ with the product topology. Let $f \: 
\Omega \to \Bbb R$ by $f(V) = V_0$ and $T \: \Omega \to \Bbb R$ by $(TV)_n 
= V_{n+1}$. Let $\mu_1, \mu_2$ be two measures on $\Omega$ under which 
$T$ is ergodic. Let $\Sigma_i$ be the essential support of the 
a.c.~spectrum of the prototypical $h_\omega$ for the process 
$(\text{\rom{supp}} (\mu_i), T, \mu_i)$. If $\text{\rom{supp}}(\mu_1)  
\subset \text{\rom{supp}}(\mu_2)$, then $\Sigma_1 \supset \Sigma_2$. 
\endproclaim

\demo{Proof} Let $\omega_i \in \text{supp}(\mu_i)$ be 
$\mu_i$-prototypical. Since $\omega_1 \in \text{supp}(\mu_2)$, 
Theorem~6.1 implies $\Sigma (h_{\omega_2}) \subset \Sigma 
(h_{\omega_1})$. \qed 
\enddemo

\remark{Remark} In a sense, Theorem~1.4 is a deterministic version of 
the Kotani support theorem, so it is not surprising that it implies 
the Kotani theorem.
\endremark

While we have stated these theorems in this section for the discrete 
case, they all extend easily to the continuum case.

\newpage
\flushpar {\bf {\S 7. The Potential $\lambda \cos (n^\beta), \; \beta > 1$}}
\vskip 0.1in

Jacobi matrices with potentials of the form $V (n) \equiv \lambda \cos 
(n^\beta)$, where $\lambda,\beta$ are real parameters with $\beta > 1$, 
had been studied numerically and heuristically by Griniasty-Fishman [\grfs] 
and Brenner-Fishman [\brfs]. The particular case $1<\beta < 2$ had been 
studied in more detail by Thouless [\thou]. The numerical evidence 
indicates that for $\beta \geq 2$, such potentials exhibit ``localization'' 
with the same Lyapunov exponents as those of random potentials (Anderson 
model) with the same coupling. The case $1<\beta < 2$ is different, and far 
less conclusive. One still expects ``localization'' away from $E=0$ (the 
center of the spectrum) and for large $\lambda$, but the Lyapunov exponents 
are smaller and seem to vanish for $E=0$ and small $\lambda$. Mathematical 
results exist for the case where $\beta$ is an integer and $\lambda$ is 
large (larger than 2, to be precise), in which case it is known [\goso,\jito] 
that there is no absolutely continuous spectrum. More precisely, for every 
polynomial $p(n)$ with a leading coefficient that is an irrational multiple 
of $\pi$, it is known that $\lambda \cos (p(n))$ can be obtained as a 
realization (an element) of an ergodic family of potentials coming from a 
suitable ergodic transformation on the $d$-dimensional torus [\cfs] (with $d$ 
the degree of $p(n)$); and that the corresponding ergodic families have only 
positive Lyapunov exponents as long as $\lambda > 2$ [\goso,\jito]. Further, 
the corresponding ergodic families are minimal [\furs], and so it follows 
from our Theorem~6.1 that {\it every} realization of such a family has no 
absolutely continuous spectrum. We note that this is also true for the case 
$\beta = 1$, where the absence of a.c.~spectrum follows from earlier results 
[\avs]. Our purpose in this section is to extend these results to cases where 
$\beta$ is not an integer. We discuss half-line problems here, and denote by 
$h_0^+$ the free Laplacian on $\ell^2 ({\Bbb Z}^+)$. The results are also 
valid for full-line problems if we replace $n$ by $|n|$. We shall prove 
the following:

\proclaim{Theorem 7.1} 
For any $\lambda >2$ and $\beta > 1$, $\Sigma_{\text{\rom{ac}}} 
(h_0^+ + \lambda \cos (n^\beta)) = \emptyset$.  
\endproclaim

\remark{Remarks} 1.  $\cos(\,\cdot\,)$ in the above theorem can be replaced 
by any real analytic function $f(\,\cdot\,)$ of period $2\pi$, in which 
case the theorem would hold for $\lambda$ ``large enough.'' This follows 
from the argument below combined with the results of Goldsheid-Sorets 
[\goso]. The explicit $\lambda > 2$  for the $\cos(\,\cdot\,)$ case is 
due to Jitomirskaya [\jito].

2. The result is actually also more general in the sense that one can 
replace $n^\beta$ by, for example, $\sum_{j=1}^k a_j n^{\beta_j}$, where 
$k$ is any positive integer, $\beta_j > 0$ for each $j$, and the $a_j$'s 
are some real numbers (except if the largest $\beta_j$ is an integer, 
in which case we would need some further condition, such as that the 
corresponding $a_j$ would be an irrational multiple of $\pi$). 

3.  For $1 < \beta < 2$, the result also follows for $\lambda = 2$, by 
results of Helffer-Sj\"ostrand [\hesj] and Last [\lastz].
\endremark

\demo{Proof} Let $\lambda > 2$. We only need to consider the case where 
$\beta$ is not an integer. Fix $k < \beta < k+1$, where $k$ is an integer, 
and consider $(n+m)^\beta$ where $n$ is large and $m\ll n$. By writing 
$(n+m)^\beta = n^\beta (1+n/m)^\beta$ and expanding $(1+n/m)^\beta$ as a 
Taylor series, we obtain:
$$
(n+m)^\beta = \sum_{\ell=0}^\infty a_\ell n^{\beta -\ell} m^\ell = 
\sum_{\ell=0}^k a_\ell n^{\beta -\ell} m^\ell + O(n^{\beta-k-1}m^{k+1}), 
\tag 7.1
$$
where $a_0 = 1$, $a_\ell = (1/\ell !)\prod_{j=0}^{\ell-1}(\beta-j)$ for 
$\ell\geq 1$. Let $b_{\ell,n} = \langle a_\ell n^{\beta -\ell}/2\pi\rangle$ 
where $\langle\,\cdot\, \rangle$ denotes fractional part (i.e., $\langle 
x\rangle = x-[x]$). Since $(n+1)^{\beta -k}-n^{\beta -k} \to 0$, 
$\{b_{k,n}\}_{n=1}^\infty$ is clearly dense in $[0,1]$ and we can pick 
a convergent subsequence $\{n_j\}$ so that $b_{k,n_j}\to b_{k,\infty}\equiv 
b_k$, where $b_k$ is an irrational. Moreover, by compactness, we can find 
a subsequence of that for which $b_{\ell,n_j}\to b_\ell$ for all $\ell\leq k$, 
where the $b_\ell$'s (for $\ell < k$) are some numbers in $[0,1]$. For the 
resulting polynomial $p(n)\equiv\sum_{\ell=0}^k b_\ell n^\ell$ we see from 
(7.1) that $\langle p(n)/2\pi\rangle$ is a pointwise limit of translations 
of $\langle n^\beta/2\pi\rangle$. Thus, the potential $\lambda\cos (p(n))$ 
is a right limit of $\lambda \cos (n^\beta)$, and by Theorem~1.4, we have 
$\Sigma_{\text{\rom{ac}}} (h_0^+ + \lambda \cos (n^\beta)) \subset 
\Sigma_{\text{\rom{ac}}} (h_0^+ + \lambda \cos (p(n))) = \emptyset$ (where 
the last equality follows from the discussion above). \qed
\enddemo

In the above proof we only needed to show that some fixed realization of 
a suitable ergodic process is obtained as a limit of translations of 
$\langle n^\beta/2\pi\rangle$. However, since the underlying ergodic systems 
are minimal, it follows that translations of that realization are themselves 
dense in the ergodic family. Thus, one sees that we can actually obtain 
{\it every} realization as such a limit. We can combine this with Kotani's 
result [\kotfmv] --- that ergodic potentials taking finitely many values have 
no absolutely continuous spectrum (unless they are periodic) --- to show that 
if $f(\,\cdot\,)$ is any real periodic piecewise constant function on the 
line (with only finitely many discontinuities per period), then for any 
$\beta > 1$ that is not an integer, $\Sigma_{\text{\rom{ac}}} 
(h_0^+ + f(n^\beta)) = \emptyset$. The proof here is very similar to 
that of Theorem~7.1, except that we need to choose a realization that 
does not take values in any of the points where $f$ is discontinuous. 

Finally, we would like to discuss the special case $1 < \beta < 2$ and 
to explain how one could prove $\Sigma_{\text{\rom{ac}}} 
(h_0^+ + \lambda \cos (n^\beta)) = \emptyset$ also for $\lambda < 2$, if 
one could prove that ``Hofstadter's butterfly has wings.'' Noting that 
we could choose the largest order coefficient $b_k$ in the proof of 
Theorem~7.1 at will, we obtain for $1 < \beta < 2$: 

\proclaim{Proposition 7.2} $\Sigma_{\text{\rom{ac}}} 
(h_0^+ + \lambda \cos (n^\beta)) \subset \cap_{\alpha\in F} 
\Sigma_{\text{\rom{ac}}} (h_0 + \lambda \cos (\pi\alpha n))$ for any 
countable set $F$ of irrational $\alpha$'s. 
\endproclaim

$\Sigma_{\text{\rom{ac}}} (h_0^+ + \lambda \cos (n^\beta )) = \emptyset$ 
would thus follow, if we could prove the following:

\example{Conjecture} Fix $\lambda \neq 0$ and $E_0$ real. Then there 
exists $\delta > 0$ and irrational $\alpha$ with $\sigma 
(h_0 + \lambda \cos (\pi\alpha n)) \cap (E_0 - \delta, E_0 + \delta) = 
\emptyset$.
\endexample

Intuitively, this conjecture comes from the fact that for $\lambda$ small, 
$h_0 + \lambda \cos (\pi\alpha n)$ should have a gap about the energy $E 
=2\cos (\pi\alpha)$. In numerical drawings of Hofstadter-like butterflies 
for various values of $\lambda$ [\ght], one indeed sees those ``stripes'' 
appear for small $\lambda$, and then broaden and get more structure as 
$\lambda$ increases, up to the critical point $\lambda = 2$ where they form 
the wings of {\it the} famous Hofstadter butterfly. Unfortunately, we do 
not know how to prove that these wings exist.

\vskip 0.3in
\flushpar {\bf {\S 8. Transfer Matrices and Bound States}}
\vskip 0.1in 

In this section, we will prove Theorem~1.7. As noted already, Theorem~1.6 
is motivation for considering $\sum^\infty_{n=1} \| T_E (n) \|^{-2}$ as an 
indicator of bound states. If it is infinite, $hu=Eu$ has no solution 
$L^2$ at infinity. 

\example{Example 1} Take $V =0$ and $E=2$. Then $hu = Eu$ has the solutions
$$
u(n) = c_1 + c_2 n, 
$$
none of which are $\ell^2$. But
$$
T_E (n) = \pmatrix n+1 & - n \\ n & 1-n \endpmatrix
$$
has $\|T_E (n)\| = \sqrt2 n + 0(1)$ and thus $\sum^\infty_{n+1} \|T_E (n) 
\|^{-2} < \infty$. We see that $\sum^\infty_{n=1} \|T_E (n) \|^{-2} \mathbreak < 
\infty$ does not imply that there is an $\ell^2$ solution. 
\endexample

\example{Example 2} Let $V(n) = c_0 n^{-2}$, $n= 1,2,\dots$ with $c_0 < 
\frac14$ and $E=2$. Then standard arguments (variation of parameters) show 
that there are two solutions $u_\pm (n)$ with
$$
u_\pm (n) \sim n^{\alpha_\pm}
$$
with $\alpha_\pm$ the roots of $\alpha (\alpha - 1) + c_0 =0$, that is,
$$
\alpha_\pm = \tfrac12 \pm \sqrt{\tfrac14 - c_0} \, .
$$
We see that $\|T(n)\| \sim C n^{\alpha_+}$, so since $\alpha_+ > \frac 12$, 
$\sum_n \|T(n)\|^{-2} < \infty$. But if $c_0 > 0$, there is no bounded 
solution. Thus, $\sum_n \|T(n)\|^{-2} < \infty$ need not even imply that 
there is a bounded solution! 
\endexample

The following (note (8.5) and (8.7)) includes the first part of Theorem~1.7 
as a special case. Its proof just abstracts Ruelle [\rue]:

\proclaim{Theorem 8.1} Let $A_1, A_2, \dots$ be unimodular $2\times 2$ real 
matrices and let $T(n) = A_n A_{n-1} \mathbreak \dots A_1$. Suppose that
$$
\sum^\infty_{n=1} \frac{\| A_{n+1}\|^2}{\| T(n)\|^2} < \infty. \tag 8.1
$$
Then there is a unit vector $u\in \Bbb R^2$ so that for any other unit 
vector $v\in \Bbb R^2$, we have
$$
\frac{\|T(n) u\|}{\|T(n) v\|} \to 0. \tag 8.2
$$
\endproclaim

\demo{Proof} Let $t(n) = \| T(n)\|$ and $a(n) = \| A_n \|$. Since $|T(n)|$ 
is self-adjoint and unimodular, it has eigenvalues $t(n)$ and $t(n)^{-1}$. 
Thus, taking $u_\theta = \binom{\cos (\theta)}{\sin (\theta)}$ we see there 
exists $\theta_n$ so that 
$$
\| T(n) u_\theta \|^2 = t(n)^2 \sin^2 (\theta - \theta_n ) + t(n)^{-2} 
\cos^2 (\theta - \theta_n) \tag 8.3
$$
for pick $\theta_n$ so that $|T(n)| u_{\theta_n} = t(n)^{-1} u_{\theta_n}$. 
Now by (8.3) for $n+1$,
$$\align
t(n+1)^2 \sin^2 (\theta_n - \theta_{n+1} ) &\leq \|T(n+1) u_{\theta_n} \|^2 \\
& \leq a (n+1)^2 \|T(n) u_{\theta_n}\|^2 \\
&= a(n+1)^2 t(n)^{-2}.
\endalign
$$

Since $A_{n+1}$ is unimodular, $t(n)=\| T(n) \| \leq \|T(n+1)\| \, 
\|A^{-1}_{n+1}\| = t(n+1) a(n+1)$ so
$$
t(n)^2 \sin^2 (\theta_n - \theta_{n+1}) \leq a(n+1)^4 t(n)^{-2}.
$$
Since $\sin^2 (x) \geq (\frac{2x}{\pi})^2$, we see that
$$
|\theta_n - \theta_{n+1}| \leq \frac{\pi}{2} \, \frac{a(n+1)^2}{t(n)^2}\, . 
\tag 8.4
$$

Thus, (8.1) implies
$$
\sum_n |\theta_n - \theta_{n+1}| < \infty.
$$
So if (8.1) holds, $\theta_n$ has a limit $\theta_\infty$ and
$$
|\theta_n - \theta_\infty | \leq \frac{\pi}{2} \sum^\infty_{m=n} \, 
\biggl[\frac{a(m+1)^2}{t(m)^2} \biggr]. \tag 8.5
$$
Let $u_\infty = u_{\theta_\infty}$ and $v_\infty = u_{\pi/2 + \theta_\infty}$. 
Since $\theta_n - \theta_\infty \to 0$, for $n$ large enough, we have by (8.3) 
that 
$$
\| T(n) v_\infty \|^2 \geq \frac12 \, t(n)^2. \tag 8.6
$$
On the other hand, by (8.3) again,
$$
\| T(n) u_\infty \|^2 \leq t(n)^2 (\theta_n - \theta_\infty)^2 + t(n)^{-2}. 
\tag 8.7
$$

(8.6) and (8.7) imply that
$$
\frac{\|T(n) u_\infty \|^2}{\|T(n) v_\infty\|^2} \leq 2 
(\theta_n - \theta_\infty )^2 + 2 t(n)^{-4} \to 0
$$
since $a(n+1)\geq 1$ and (8.1) imply that $t(n)\to\infty$. From this, (8.2) 
follows. \qed
\enddemo

The following includes the second part of Theorem~1.7 as a special case 
(where $a(n)$ is bounded):

\proclaim{Theorem 8.2} Under the hypothesis of Theorem~{\rom{8.1}}, 
suppose that we also have that 
$$
\sum^\infty_{m=1} \|T(m)\|^2 \biggl( \, \sum^\infty_{n=m} 
\frac{\| A (n+1)\|^2}{\|T(n)\|^2} \biggr)^2 <\infty. \tag 8.8
$$
Then there is a unit vector $u_\infty$ with
$$
\sum^\infty_{n=1} \|T(n) u_\infty \|^2 < \infty.
$$
\endproclaim

\demo{Proof} Using (8.7) and (8.5), we see that $\sum^\infty_{n=1} \| T(n) 
u_\infty \|^2 <\infty$ if (8.8) holds and if $\sum_n t(n)^{-2} < \infty$. 
But since $a(n) \geq 1$, (8.1) implies that $\sum_n t(n)^{-2}<\infty$. \qed
\enddemo

\example{Example 3} Suppose that $\| A(n)\|$ is bounded and $t(n) \sim 
n^\gamma$ in the sense that $C_- n^\gamma \leq t(n) \leq C_+ n^\gamma$. 
Then (8.1) requires $\gamma > \frac12$ while (8.4) requires $\gamma > 
\frac 32$. Notice in Example~2, $\gamma = \alpha_+$ while an $\ell^2$ 
solution requires $\alpha_- < -\frac12$. Since $\alpha_+ = 1 - \alpha_-$, 
Example~2 provides an example with $\gamma = \frac32$ where there is no 
$\ell^2$ solution (namely, take $c_0 = -\frac34$). Thus, $\gamma > \frac32$ 
is best possible!
\endexample

If one has control over the limit of $\ln \|T(n)\|$, one can say more:

\proclaim{Theorem 8.3} Suppose that the hypotheses of Theorem~{\rom{8.1}} 
hold and that
$$
\lim\limits_{n\to\infty}\, \frac{\ln \|T(n)\|}{f(n)} = 1
$$
and
$$
\lim\limits_{n\to\infty}\, \frac{\ln \|A(n)\|}{f(n)} = 0, 
$$
where $f(n)$, a monotone increasing function, is such that
$$
\sum e^{-\epsilon f(n)} < \infty
$$
for any $\epsilon >0$. Then the $u_\infty$ of Theorem~{\rom{8.2}} obeys
$$
\lim\limits_{n\to\infty}\, \frac{\ln \|T(n) u_\infty\|}{f(n)} = -1.
$$
\endproclaim

\demo{Proof} By (8.5) and (8.7), for any $\epsilon >0$, for $n$ large
$$\align
\|T(n) u_\infty\|^2 &\leq e^{-2(1-\epsilon)f(n)} + 
\biggl(\frac{\pi}{2}\biggr)^2 \biggl( \, \sum^\infty_{m=n} 
\frac{e^{2\epsilon f(m)}}{e^{2(1-\epsilon)f(m)}} \biggr)^2 
e^{2(1+\epsilon)f(n)} \\
&\leq e^{-2(1-\epsilon)f(n)} + \biggl(\frac{\pi}{2}\biggr)^2 
e^{-(2-11\epsilon)f(n)} \biggl[ \, \sum^\infty_{m=n} 
e^{-\epsilon f(m)} \biggr]
\endalign
$$
so
$$
\varlimsup \, \frac{\|T(n) u_\infty \|}{f(n)} \leq -1.
$$
On the other hand, $\| T(n) u\|_\infty \geq \|T(n) \|^{-1}$ implies
$$
\varliminf \, \frac{\|T(n) u_\infty \|}{f(n)} \geq -1. \qed
$$
\enddemo

Typical cases of this theorem are $f(n) = n^\alpha$; $f(n)=n$ is Ruelle's 
theorem. For the case $f(n)=\ln (n)$ where (8.9) fails, we have

\proclaim{Theorem 8.4} Suppose the hypotheses of Theorem~{\rom{8.1}} hold 
and that 
$$\align
\lim\limits_{n\to\infty}\, \frac{\ln \|T(n)\|}{\ln n} &= \gamma \\
\lim\limits_{n\to\infty}\, \frac{\ln \|A(n)\|}{\ln n} &= 0
\endalign
$$
where $\gamma > \frac12$. Then
$$
\varlimsup\limits_{n\to\infty}\, \frac{\ln \|T(n) u_\infty\|}{\ln n} 
\leq 1-\gamma \tag 8.10
$$
while
$$
\varliminf\limits_{n\to\infty}\, \frac{\ln \|T(n) u_\infty\|}{\ln n} 
\geq -\gamma.  \tag 8.11
$$
\endproclaim

\demo{Proof} As in the last theorem, (8.11) is a consequence of 
$\|T(n) u_\infty \| \geq \| T(n)\|^{-1}$. To get (8.10), we use (8.5), 
(8.7) to see that for any $\epsilon > 0$, 
$$\align
\| T(n) u_\infty \|^2 &\leq n^{-2\gamma +\epsilon} + 
\biggl( \, \sum^\infty_{m=n} m^{-2\gamma + \epsilon} \biggr)^2 
n^{2\gamma + \epsilon} \\
&\leq n^{-2\gamma + \epsilon} + Cn^{2-2\gamma + 2\epsilon}. \qed
\endalign
$$
\enddemo

Example~2 shows there are cases where the limit is $1-\gamma$. [\kls] 
has examples where the limit is $-\gamma$.

The ideas of this section can be applied to certain continuum problems 
by sampling the wave function at a discrete set of points.

\vskip 0.3in
\flushpar {\bf {Appendix: BGK Eigenfunction Expansions}}
\vskip 0.1in

The proofs of the estimates in Section~2 are one-dimensional, relying 
on the relation between Green's functions and $m$-functions. Our goal 
in this appendix is to discuss an alternate proof which extends to 
higher dimensions. The applications of these estimates in Section~3 
are intrinsically one-dimensional, so those results do not extend to 
higher dimensions. Nevertheless, we believe these general estimates 
may be of use elsewhere. 

We recall the abstract eigenfunction expansion dubbed BGK expansions in 
[\ssg] after work of Berezinski, Browden, Garding, Gel'fand, and Kac 
(see [\ssg] for references).  In the discrete case, they take the 
following form. Let $V$ be an arbitrary function on $\Bbb Z^\nu$. 
Let $(Hu)(n) =\sum_{|\ell| = 1} u(n + \ell) + V(n) u(n)$. Then there 
exist measures $\{d\rho_k (E)\}^\infty_{k=1}$ on $\Bbb R$ and a measure 
$d\rho_\infty (E)$ so that the $d\rho$'s are mutually singular 
[\aron,\dono,\svan]. Moreover, for a.e.~$E$ w.r.t.~$d\rho_k (E)$, 
there exist $k$ linearly independent functions $\{ u_{j,k} 
(n; E)\}^k_{j=1}$ on $\Bbb Z^\nu$ so that
\roster
\item"{(i)}" $\sum_{|\ell| =1} u_j (n+\ell) + (V(n)-E) u_j(n) = 0$. 
\item"{(ii)}" For any $f$ on $\Bbb Z^\nu$ of finite support, define 
$a_{j,k} (f)(E) = \sum_n \overline{u_{j,k} (n; E)}\, f(n)$. Then
$$
a_{j,k} (Hf)(E) = E a_{j,k}(f)(E)
$$
and for any $f,g$ of finite support,
$$
\langle f,g\rangle = \sum^{\text{``$\infty$"}}_{k=1} \, \sum^k_{j=1} \int 
\overline{a_{j,k} (f)(E)} \, a_{j,k} (g)(E) \, d\rho_k (E) \tag A.1
$$
with an explicit $\rho_\infty$-term intended in 
$\sum^{\text{``$\infty$"}}_{k=1}$.
\endroster

Pick $f=g=\delta_n$, a Kronecker delta function at $n$. Then (A.1) becomes
$$
\sum^{\text{``$\infty$"}}_{k=1}\, \sum^k_{j=1} \int |u_{j,k} (n;E)|^2 \, 
d\rho_k (E) =1.\tag A.2
$$

This is essentially (2.6D) except in arbitrary dimension. In the 
one-dimensional case, $d\rho_k =0$ for $k\neq 1$. If we define $d\tilde\rho 
(E) = |u(1; E)|^2\, d\rho_1$ and $\tilde u (n; E) = u(n; E) u(1;E)^{-1}$, 
then (A.2) is exactly (2.6D).

In the continuum case, the situation is similar. One needs some minimal 
local regularity on $V$ (see [\ssg]). Using the fact that $e^{-tH} (x,x) = 
0(t^{-1/2})$ as $t\downarrow 0$, one can show that as $f\to\delta_x$, a 
$\delta$-function at $x$, $(f, (H+c)^{-\ell}f)$ stays finite and bounded 
in $x$ so long as $2\ell > \nu$. Thus, (A.2) in the continuum case becomes 
$$
\sum^{\text{``$\infty$"}}_{k=1} \, \sum^k_{j=1} \int |u_{j,k} (x; E)|^2 
\frac{d\rho_k (E)}{(1+|E|)^\ell} \leq C
$$
uniformly in $x$ where $\ell > \frac{\nu}2$.

As in Section~3, from these bounds and Fatou's lemma, we get bounds like 
before for a.e.~$E$ w.r.t.~$\sum_k d\rho_k (E)$, 
$$
\varliminf\limits_{L\to\infty} \, \frac{1}{(2L+1)^\nu} 
\sum_{|n| \leq L} |u(n; E)|^2 < \infty.
$$

\vskip 0.3in
\Refs
\endRefs
\vskip 0.1in
\item{\aron.}\ref{N.~Aronszajn}{On a problem of Weyl in the theory of 
Sturm-Liouville equations}{Am. J.~Math.}{79}{1957}{597--610}
\gap
\item{\avs.}\ref{J.~Avron and B.~Simon}{Almost periodic Schr\"odinger operators, 
II. The integrated density of states}{Duke Math.~ J.}{50}{1983}{369--391}
\gap
\item{\brfs.}\ref{N.~Brenner and S.~Fishman}{Pseudo-randomness and localization}
{Nonlinearity}{4}{1992}{211--235}
\gap
\item{\car.}\ref{R.~Carmona}{One-dimensional Schr\"odinger operators with random 
or deterministic potentials, New spectral types}{J.~Funct.~Anal.}{51}{1983}
{229--258}
\gap
\item{\cala.} R.~Carmona and J.~Lacroix, {\it{Spectral Theory of Random 
Schr\"odinger Operators}}, Birk-h\"auser, Boston, 1990.
\gap
\item{\cfs.} I.P.~Cornfeld, S.V.~Fomin, and Ya.G.~Sinai, {\it{Ergodic Theory}},
Springer, New York-Heidelberg-Berlin, 1982.
\gap
\item{\cyc.} H.L.~Cycon, R.G.~Froese, W.~Kirsch, and B.~Simon, 
{\it{Schr\"odinger Operators}}, Springer, Berlin-Heidelberg-New York, 1987.
\gap
\item{\dasi.}\ref{E.B.~Davies and B.~Simon}{Scattering theory for systems with 
different spatial asymptotics on the left and right}{Commun.~Math.~Phys.}{63}
{1978}{277--301}
\gap
\item{\dasii.}\ref{E.B.~Davies and B.~Simon}{$L^1$-properties of intrinsic 
Schr\"odinger semigroups}{J.~Funct. Anal.}{65}{1986}{126--146}
\gap
\item{\deis.}\ref{P.~Deift and B.~Simon}{Almost periodic Schro\"odinger 
operators, III. The absolutely continuous spectrum in one dimension}
{Commun.~Math.~Phys.}{90}{1983}{389--411}
\gap
\item{\del.}\ref{F.~Delyon}{Apparition of purely singular continuous spectrum 
in a class of random Schr\"odinger operators}{J.~Statist.~Phys.}{40}{1985}
{621--630}
\gap
\item{\dss.}\ref{F.~Delyon, B.~Simon, and B.~Souillard}{From power pure point 
to continuous spectrum in disordered systems}{Ann.~Inst.~H.~Poincar\'e}{42}
{1985}{283--309}
\gap
\item{\dono.}\ref{W.~Donoghue}{On the perturbation of the spectra}{Commun.~Pure 
Appl.~Math.}{18}{1965}{559--579}
\gap
\item{\furs.}\ref{H.~Furstenberg}{Strict ergodicity and transformations of the 
torus}{Am.~ J.~Math.}{83}{1961}{573--601}
\gap
\item{\gp.}\ref{D.J.~Gilbert and D.~Pearson}{On subordinacy and analysis of the 
spectrum of one-dimensional Schr\"odinger operators}{J.~Math.~Anal.}{128}
{1987}{30--56}
\gap
\item{\goso.}\ref{I.~Goldsheid and E.~Sorets}{Lyapunov exponents of the 
Schr\"odinger equation with certain classes of ergodic potentials}
{Amer.~Math.~Soc.~Trans. (2)}{171}{1996}{73--80}
\gap
\item{\gor.}\ref{A.Ya.~Gordon}{Deterministic potential with a pure point 
spectrum}{Math.~Notes}{48}{1990}{1197--1203}
\gap
\item{\grfs.}\ref{M.~Griniasty and S.~Fishman}{Localization by pseudorandom 
potentials in one dimension}{Phys.~Rev.~Lett.}{60}{1988}{1334--1337}
\gap
\item{\ght.}\ref{J.P.~Guillement, B.~Helffer, and P.~ Treton}
{Walk inside Hofstadter's butterfly}{J.~Phys. France}{50}{1989}{2019--2058}
\gap
\item{\hesj.}\ref{B.~Helffer and J.~Sj\"ostrand}{Semi-classical analysis for 
Harper's equation, III. Cantor structure of the spectrum}
{M\'em.~Soc.~Math.~France (N.S.)}{39}{1989}{1--139}
\gap
\item{\ish.}\ref{K.~Ishii}{Localization of eigenstates and transport 
phenomena in one-dimensional disordered systems}{Suppl.~Prog.~Theor.~Phys.}
{53}{1973}{77--138}
\gap
\item{\jito.} S.~Jitomirskaya, to be published.
\gap
\item{\jlprl.}\ref{S.~Jitomirskaya and Y.~Last}{Dimensional Hausdorff properties
of singular continuous spectra}{Phys.~Rev.~Lett.}{76}{1996}{1765--1769}
\gap
\item{\jli.} S.~Jitomirskaya and Y.~Last, {\it{Power law subordinacy and singular
spectra, I. Half-line operators}}, in preparation.
\gap
\item{\js.}\ref{S.~Jitomirskaya and B.~Simon}{Operators with singular continuous 
spectrum, III. Almost periodic Schr\"odinger operators}{Commun.~Math.~Phys.}
{165}{1994}{201--205}
\gap
\item{\kap.}\ref{S.~Kahn and D.B.~Pearson}{Subordinacy and spectral theory for
infinite matrices}{Helv. Phys.~Acta}{65}{1992}{505--527}
\gap
\item{\katz.} Y.~Katznelson, {\it{An Introduction to Harmonic Analysis}},
Dover, New York, 1976.
\gap
\item{\kis.}\ref{A.A.~Kiselev}{Absolutely continuous spectrum of one-dimensional 
Schr\"odinger operators and Jacobi matrices with slowly decreasing potentials}
{Commun.~Math.~Phys.}{179}{1996}{377--400}
\gap
\item{\kls.} A.~Kiselev, Y.~Last, and B.~Simon, {\it{Modifed Pr\"ufer and EFGP 
transforms and the spectral analysis of one-dimensional Schr\"odinger operators}}, 
to appear in Commun.~Math. Phys.
\gap
\item{\kot.} S.~Kotani, {\it{Ljapunov indices determine absolutely continuous 
spectra of stationary random one-dimensional Schr\"odinger operators}}, 
Stochastic Analysis (K.~Ito, ed.), pp.~225--248, North Holland, Amsterdam, 1984.
\gap
\item{\kots.}\ref{S.~Kotani}{Support theorems for random Schr\"odinger operators}
{Commun.~Math.~Phys.}{97}{1985}{443--452}
\gap
\item{\kotfmv.}\ref{S.~Kotani}{Jacobi matrices with random potential taking 
finitely many values}{Rev.~Math. Phys.}{1}{1989}{129--133}
\gap
\item{\lastz.}\ref{Y.~Last}{Zero measure spectrum for the almost Mathieu operator}
{Commun.~Math.~Phys.}{164}{1994}{421--432}
\gap
\item{\lath.} Y.~Last, {\it{Conductance and Spectral Properties}}, D.Sc.~thesis,
Technion, 1994.
\gap
\item{\lapaii.} Y.~Last, {\it{Periodic approximants of Jacobi matrices, II. 
Complete determination of the absolutely continuous spectrum}}, in 
preparation.
\gap
\item{\lsii.} Y.~Last and B.~Simon, {\it{Modified Pr\"ufer and EFGP transforms 
and deterministic models with dense point spectrum}}, to appear in 
J.~Funct.~Anal.
\gap
\item{\lesa.} B.M.~Levitan and I.S.~Sargsjan {\it{Introduction to Spectral 
Theory}}, Trans.~Math.~Monographs, {\bf 39}, Amer.~Math.~Soc., Providence, RI, 
1976.
\gap
\item{\pas.}\ref{L.~Pastur}{Spectral properties of disordered systems in the 
one-body approximation}{Commun.~Math.~Phys.}{75}{1980}{167--196}
\gap
\item{\pcmp.}\ref{D.~Pearson}{Singular continuous measures in scattering theory}
{Commun.~Math.~Phys.}{60}{1978}{13--36}
\gap
\item{\pea.} D.~Pearson, {\it{Pathological spectral properties}}, Mathematical 
Problems in Theoretical Physics, pp.~49--51, Lecture Notes in Physics No.~80, 
Springer, Berlin-New York, 1979.
\gap
\item{\rue.}\ref{D.~Ruelle}{Ergodic theory of differentiable dynamical systems}
{Publ.~Math.~IHES}{50}{1979}{275--306}
\gap
\item{\sch.} I.~Sch'nol, {\it{On the behavior of the Schr\"odinger equation}},  
Mat.~Sb. {\bf 42} (1957), 273--286 [in Russian].
\gap
\item{\ssg.}\ref{B.~Simon}{Schr\"odinger semigroups}{Bull.~Amer.~Math.~Soc.}
{7}{1982}{447--526}
\gap 
\item{\scmp.}\ref{B.~Simon}{Some Jacobi matrices with decaying potentials and 
dense point spectrum}{Commun.~Math.~Phys.}{87}{1982}{253--258}
\gap
\item{\svan.} B.~Simon, {\it{Spectral analysis and rank one perturbations and 
applications}}, CRM Lecture Notes Vol. 8 (J.~Feldman, R.~Froese, L.~Rosen, eds.), 
pp.~109--149, Amer.~Math.~Soc., Providence, RI, 1995.
\gap
\item{\ssp.} \ref{B.~Simon and T.~Spencer}{Trace class perturbations and the 
absence of absolutely continuous spectrum}{Commun.~Math.~Phys.}{125} 
{189}{113--126}
\gap
\item{\sst.} \ref{B.~Simon and G.~Stolz}{Operators with singular continuous 
spectrum, V. Sparse potentials}{Proc.~Amer.~Math.~Soc.}{124}{1996}{2073--2080}
\gap
\item{\thou.}\ref{D.J.~Thouless}{Localization by a potential with slowly 
varying period}{Phys.~Rev.~Lett.}{61}{1988}{2141--2143}
\gap
\enddocument